\documentclass[nonacm]{acmart} 

\usepackage{amsmath,amsfonts,bbm}
\usepackage{algorithmic}
\usepackage{graphicx}
\usepackage{textcomp}
\usepackage{xcolor}
\usepackage{multirow}
\usepackage{makecell}
\def\BibTeX{{\rm B\kern-.05em{\sc i\kern-.025em b}\kern-.08em
    T\kern-.1667em\lower.7ex\hbox{E}\kern-.125emX}}
\usepackage{setspace}
\usepackage{fancyhdr}
\usepackage{kantlipsum}
\fancypagestyle{plain}{
\fancyhf{}
\fancyhead[C]{Conference on \LaTeX} 
} \usepackage{eso-pic}

\AtBeginDocument{%
  \providecommand\BibTeX{{%
    Bib\TeX}}}

\setcopyright{acmlicensed}
\copyrightyear{2023}
\acmYear{2023}
\acmDOI{XXXXXXX.XXXXXXX}

\acmJournal{TWEB}
\acmVolume{0}
\acmNumber{0}
\acmArticle{0}
\acmMonth{0}

\begin{document}

\title{The Pulse of Mood Online: Unveiling Emotional Reactions in a Dynamic Social Media Landscape}


\author{Siyi Guo}
\affiliation{%
  \institution{Information Sciences Institute}
  \city{Marina del Rey}
  \country{USA}}
\email{siyiguo@usc.edu}

\author{Zihao He}
\affiliation{%
  \institution{Information Sciences Institute}
  \city{Marina del Rey}
  \country{USA}}
\email{zihaoh@usc.edu}

\author{Ashwin Rao}
\affiliation{%
  \institution{Information Sciences Institute}
  \city{Marina del Rey}
  \country{USA}}
\email{mohanrao@usc.edu}

\author{Fred Morstatter}
\affiliation{%
  \institution{Information Sciences Institute}
  \city{Marina del Rey}
  \country{USA}}
\email{fredmors@isi.edu}

\author{Jeffrey Brantingham}
\affiliation{%
  \institution{University of California, Los Angeles}
  \city{Los Angeles}
  \country{USA}}
\email{branting@ucla.edu}

\author{Kristina Lerman}
\affiliation{%
  \institution{Information Sciences Institute}
  \city{Marina del Rey}
  \country{USA}}
\email{lerman@isi.edu}

\renewcommand{\shortauthors}{Guo et al.}

\begin{abstract}
The rich and dynamic information environment of social media provides researchers, policy makers, and entrepreneurs with opportunities to learn about social phenomena in a timely manner. However, using these data to understand social behavior is difficult due to heterogeneity of topics and events discussed in the highly dynamic online information environment. To address these challenges, we present a method for systematically detecting and measuring emotional reactions to offline events using change point detection on the time series of collective affect, and further explaining these reactions using a transformer-based topic model. We demonstrate the utility of the method by successfully detecting major and smaller events on three different datasets, including (1) a Los Angeles Tweet dataset between Jan. and Aug. 2020, in which we revealed the complex psychological impact of the BlackLivesMatter movement and the COVID-19 pandemic, (2) a dataset related to abortion rights discussions in USA, in which we uncovered the strong emotional reactions to the overturn of Roe v. Wade and state abortion bans, and (3) a dataset about the 2022 French presidential election, in which we discovered the emotional and moral shift from positive before voting to fear and criticism after voting. The capability of our method allows for better sensing and monitoring of population's reactions during crises using online data.
\end{abstract}




\maketitle

\section{Introduction}
Social media platforms connect billions of people worldwide, enabling them to exchange information and opinions, express emotions,  
and to respond to others. 
Researchers, policy makers, and entrepreneurs have grown interested in learning what the unfettered exchange of information reveals about current social conditions, 
including using social media data to track public opinion on  important issues~\cite{Barbera2015measuring, he2022infusing} 
and monitor the well-being of populations at an unprecedented spatial scale and temporal resolution~\cite{pellert2022validating}. 

Using social media data to learn about human behavior, however, poses significant challenges. Social media represents a heterogeneous, highly dynamic information environment where some topics are widely discussed while others are barely mentioned \cite{dodds2022fame}. It includes people's self-reports of their own lives, as well as reactions to external events. Researchers have developed methods to 
detect events from online discussions \cite{malik2022performance,weng2011event,morabia-etal-2019-sedtwik,rezaei2022event}. 
However, social media data provides evidence for learning about human behavior beyond shifts in topics. For example, it can also shed light on emotions and morality, which are important drivers of individual attitudes, beliefs, and psychological and social well-being~\cite{vanKleef2016Editorial,haidt2007moral}.

To study the collective affect, researchers investigated how social media content influences emotional user engagement~\cite{babac2022emotion,aldous2022measuring}. 
These works, however, leave a gap in our understanding of collective emotional and moral reactions to socio-political events, which delineate opinion dynamics and emergence of polarization, and help identify online influence campaigns.
%

To bridge these gaps, we present a methodology for detecting, measuring and explaining the collective emotional reactions to offline events. Using state-of-the-art transformer-based models, we construct the time series of aggregate affect from social media posts. We detect emotional reactions as discontinuities in these time series, and then explain the reactions using topic modeling.
We demonstrate the utility of the methodology on three different datasets: 
\begin{enumerate}
    \item \textit{2020 Los Angeles Tweets}: a dataset collected between Jan. and Aug. 2020, of which time span represents a complex period in American history with important social, political and cultural changes. We successfully detect the simultaneous crises of the COVID-19 pandemic and racial justice reckoning, and other important events like political primaries. We show how these developments had profound impact on the psychological state of the population.
    \item \textit{2022 Abortion Tweets}: a tweet collection related to abortion rights and the overturning of Roe v. Wade, spanning the entire year of 2022. We uncovered various abortion related events and discussions, such as the leak of SCOTUS ruling, the overturn of Roe v. Wade, state abortion bans, and the complex emotional reactions to the 2022 US Midterm election.
    \item \textit{2022 French Election Tweets}: a dataset about the 2022 French presidential election, of which time coincided with other major events, the Russia-Ukraine war and the G7 summit. We detect the correct dates of the first and second rounds of the presidential and legislative elections, as well as the shift in population's emotional responses from positive before each voting round to negative afterwards.
\end{enumerate}

The success of our method on three different datasets indicates its effectiveness and generalizability. We show several benefits of our emotional reaction detection method. First, the method is able to not only detect major events but also their complex and multifaceted emotional and moral impact. Second, this unsupervised method is also able to discover smaller events that are easily missed on news media. Third, the method can identify events that happened closely in time or even on the same day based on different emotional reactions. In addition, we also demonstrate the importance of disaggregating by topics
when studying specific issues, using the analysis on emotional reactions in sub-topics during COVID-19 pandemic as an example. Although we apply our method to Twitter datasets in this study, it is also generalizable to other social media platforms and news. Our results suggest that studying the collective emotional reactions on social media can provide valuable insights into understanding people's opinions and responses to timely socio-political events, and aid policy makers in crafting messages that align with the values and concerns of the population \footnote{Short version of this paper is published as \cite{guo2023measure}}.

\section{Related Works}

\paragraph{Event Detection:}
With the rich and dynamic information on social media that is tightly related to offline events, researchers have developed methods for event detection on online platforms, including topic detection techniques such as Latent Dirichlet Allocation (LDA) \cite{blei2003latent} and Topic2Vec \cite{niu2015topic2vec}, clustering documents based on their textual similarity \cite{malik2022performance}, studying term co-occurrence and performing term frequency analysis \cite{weng2011event} and detecting bursty terms \cite{malik2022performance,morabia-etal-2019-sedtwik}. Recent methods also incorporate deep learning techniques \cite{rezaei2022event,cao2021knowledge}. 
Although these methods help detect events from social media data, we also want to understand the dynamics of emotions and moral sentiments in the aggregate online population, as opinions and emotions expressed online have a complex interplay that can impact and manifest in offline behaviors.

\paragraph{Sentiments and Emotions:}
Early research on quantifying online emotions relied on dictionary-based approaches to measure sentiment of messages by counting the occurrences of positive or negative words~\cite{golder2011diurnal,sandoval2020sentiment}. Researchers found that the sentiment of tweets in aggregate revealed hourly, diurnal and weekly patterns of mood variations~\cite{golder2011diurnal,dodds2011temporal}. Some studies demonstrated the feasibility of monitoring subjective wellbeing of populations~\cite{jaidka2020estimating} at unprecedented temporal scale and  resolution~\cite{mitchell2013geography}. Other works used social media sentiment analysis to study user reactions to political campaigns \cite{sandoval2020sentiment}, or as an alternative to costly public opinion polls~\cite{cody2015climate}, predict stock market prices~\cite{bollen2011twitter} and results of elections~\cite{tumasjan2010predicting}.

In terms of studying emotional reactions to offline events, Hauthal et al. \cite{hauthal2019analyzing} used emojis to analyze the online reaction to Brexit. A recent work conducted emotion analysis of user reactions to online news \cite{babac2022emotion}, focusing on the relationship among news content, emotions and user engagement reactions. Another study investigates the user reactions to news articles by predicting emotional reactions before and after publishing the posts \cite{aldous2022measuring}. These works utilize user reactions such as comments, likes and shares to specific news articles, and focus more on how social media content and news articles influence emotional engagement online.

Different from aforementioned studies, we are interested in understanding the online dynamics of emotions and moral sentiments in response to the continuous stream of real-life events, because affect is tightly related to opinions and induces offline actions \cite{mirlohi2021causal}, providing us valuable information to study human behavior. We incorporate both event detection and emotion analysis techniques to detect, measure and explain emotional and moral reactions on social media.

\section{Methods and Materials}
\begin{figure}[h]
\centering
\includegraphics[width=0.98\linewidth]{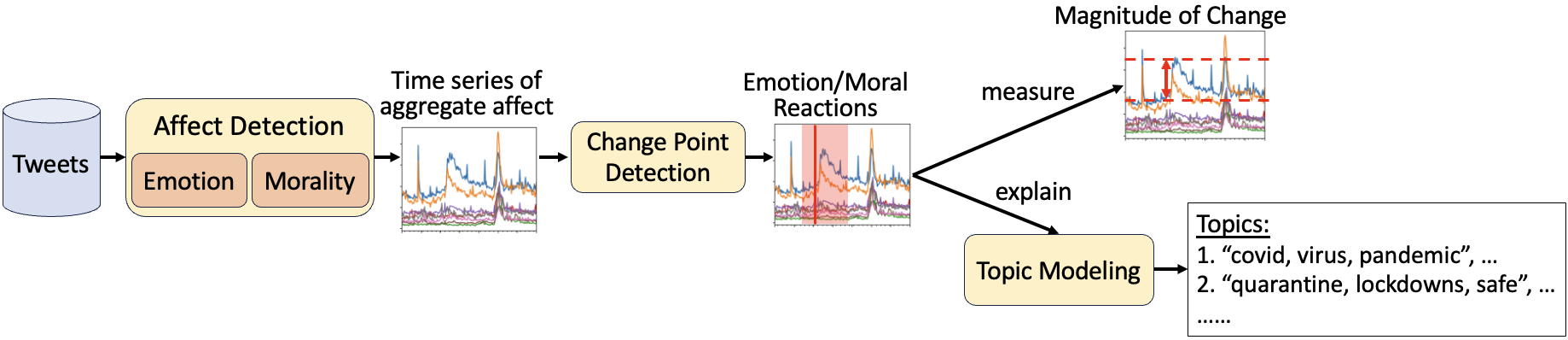} 
\caption{Pipeline to detect and measure online emotional reactions.}
\label{fig_flowchart}
\end{figure}

To understand the dynamics of affect, we propose a pipeline (Fig.~\ref{fig_flowchart}) that detects, measures and explains online emotional and moral reactions to offline events. With a set of timestamped texts, e.g. tweets, we first perform emotion and morality detection from text. We then 
construct the time series of the aggregate affect on a daily basis. Next, to detect reactions, 
we perform change point detection on each emotion and morality time series. We measure the magnitude of the change at each detected change point and perform topic modeling to explain the offline event that triggered the specific online reaction.

\subsection{Data}
\subsubsection{2020 Los Angeles Tweets}
The data was collected using Twitter's Filter API by specifying a geographic bounding box over the region of Los Angeles. 
This method collects every tweet that is either geotagged within the bounding box (using the device's coordinates with the user's permission), or by using the Twitter ``place'' feature, where the user tags their location. 
We collected 17M tweets from 350K unique users. The dataset includes a wider range of topics and authors than data collected using keywords typically do.

\subsubsection{2022 Abortion Tweets}
The data comprises tweets about abortion rights in the U.S. and the overturning of Roe vs Wade~\cite{chang2023roeoverturned}. We filter English tweets posted within the U.S. over the period of an entire year of 2022. Each tweet contains at least one term from a list of keywords that reflect both sides of the abortion debate, such as ``roevswade'', ``prochoice'', ``MyBodyMyChoice'', ``prolife'', and ``ValueLife''. The data include 12M tweets generated by 1M users.

\subsubsection{2022 French Election Tweets}
This is a corpus of 6M tweets about the 2022 French presidential election. The tweets were collected by querying Twitter with a set of keywords related to the election: e.g., ``election'', ``élection'', ``l'élection'', ``Elysee 2022'', ``Elysee2022'', etc. 90\% of the tweets were in French, and the rest were in English and other languages. 

\subsection{Emotion and Morality Detection}
We first measure emotions and moral sentiments expressed in an individual tweet. For emotions, we use a transformer-based language model SpanEmo~\cite{alhuzali-ananiadou-2021-spanemo}, fine-tuned on the SemEval 2018 1e-c data \cite{mohammad-etal-2018-semeval}. This model outperforms prior methods by learning the correlations among the emotions. 
It measures 
\textit{anticipation}, \textit{joy}, \textit{love}, \textit{trust}, \textit{optimism}, \textit{anger}, \textit{disgust}, \textit{fear}, \textit{sadness}, \textit{pessimism} and \textit{surprise}.

We quantify the moral sentiments of tweets along five dimensions~\cite{haidt2007moral}: dislike of suffering (\textit{care}/\textit{harm}), dislike of cheating (\textit{fairness}/\textit{cheating}), group loyalty (\textit{loyalty}/\textit{betrayal}), respect of authority and tradition (\textit{authority}/\textit{subversion}), and concerns with purity and contamination (\textit{purity}/\textit{degradation}). 
We fine-tune a transformer-based model on diverse training data (see \cite{guo2023data} for details). 
The large amount and the variety of topics in our training data helps mitigate the data distribution shift during inference. For both emotion and morality detection, we use the multilingual XLM-T \cite{xlm2021barbieri} as our base transformer model on the 2022 French Election data. For other English datasets, we use ``bert-base-uncased'' as the base model. After labeling tweets, we calculate the daily fractions of tweets with different emotion and moral categories to construct the time series.

\paragraph{Evaluation} We compare our emotion and morality detection methods with widely used dictionary-based methods, namely keyword matching using Emolex (does not include ``love'' category) for emotions, and Distributed Dictionary Representations (DDR) \cite{8b2871f503a14011ae81e6ab1664a638} for morality. On the Los Angeles dataset, our methods outperforms baselines on ten out of 11 emotion categories (the F1-scores of our method are in $0.42 \pm 0.14$, and those of Emolex are in $0.15 \pm 0.11$) and we outperform on nine out of ten moral categories (the F1-scores of our method are in $0.31 \pm 0.17$, and those of DDR are in $0.17 \pm 0.15$). See $\S$~\ref{sec:appendix_eval} in Appendices for details about human annotation and F1-scores. On 2022 French Election Tweets, similar evaluation was performed based on human annotation. The emotions had an average F1-score of 0.66, and the moral categories had an average F1-score of 0.58. On the 2022 Abortion data, validation was given by \citeauthor{rao2023tracking}. We notice that the model performance inevitably varies with support for different categories, as also observed in previous studies \cite{Hoover2020moral,trager2022moral}. Despite some variation in model performance, prior research~\cite{pellert2022validating} has validated that when aggregating on the collective level, the time series of sentiments constructed with supervised deep learning detection have strong correlations with those from self-reports.

\subsection{Detecting and Measuring Changes} 
\subsubsection{Change point detection}
The time series of emotions and morality reveal the complex dynamics of aggregate affect on social media. We define an emotional reaction as a change in the corresponding time series. To detect such change points, we combine two popular methods. The first, cumulative sum (CUSUM) method~\cite{hinkley1971inference}, detects a shift of means, and is good at detecting changes like the COVID-19 outbreak, which shifted the baseline emotion and moral sentiment. To detect multiple change points, we use a sliding window to scan the whole time series. We set the window size to be four weeks and slide it every three days for the best precision. Another type of event, such as Valentine's Day, creates a short surge of emotions, can be better detected with Bayesian Online Change Point Detection (BOCPD)~\cite{adams2007bayesian}. It uses Bayesian inference to determine if the next data point is improbable, which is good at detecting sudden changes. We identify a change point to be significant when either CUSUM or BOCPD gives a significant confidence score, using $0.5$ as threshold. We perform change point detection separately for each time series of emotion and morality, because different types of events may elicit different reactions.  

\subsubsection{Interrupted time series analysis}
For each detected change point, we quantify the magnitude of the collective reaction in two aspects, the short-term and the long-term changes. To calculate the short-term change, we use interrupted time series analysis \cite{penfold2013use, schaffer2021interrupted}. On the time series of daily fraction of each emotional and moral category, we select the time window from seven days before the event to three days after, as discussions on Twitter usually die down within a short time~\cite{leskovec2009meme}. We perform linear regression for each change point on each time series:
\[y = \beta_0 + \beta_1t + \beta_2\mathbbm{1}_{after\ event} + \beta_3t * \mathbbm{1}_{after\ event},\]
where $y$ is the daily fraction of an emotion/morality category, $t$ is the time variable, and $\mathbbm{1}_{after\ event}$ is a binary indicator which equals 0 before a change point and 1 after the change point. To represent the change associated with an event, we use $\frac{\beta_3}{mean(y)}$, the coefficient for the interaction term normalized by the mean of this segment of time series, and convert it to percentage. We also report the p-value from the regression along with the change.

To measure the long-term change, we compute the baseline level as the mean of the time series over seven days before the change point. Then we compare the baseline to the time series value two weeks after the event (we take a five-day average around the two-week mark). The size of the window is empirically chosen to be two weeks so that enough observations are made, but it would not be affected by another event earlier or later. We report the accompanying p-value from Student's T-test.


\subsection{Explaining Changes with Topic Modeling} \label{section_topic_modeling}
We try to explain changes in emotions detected by our method using topic modeling. 
We choose BERTopic~\cite{grootendorst2022bertopic}, a transformer-based language model that extracts highly coherent topics compared to traditional LDA. We evaluate both methods on a set of 10\% randomly selected tweets from our data, using a different numbers of topics ranging from 10 to 50 in steps of 10. Over different runs, BERTopic gives higher NPMI 
coherence scores ($0.14 \pm 0.01$) compared to LDA ($0.03 \pm 0.01$), and similar diversity \cite{dieng-etal-2020-topic} scores ($0.75 \pm 0.04$) compared to LDA ($0.76 \pm 0.04$).

For each emotional reaction, we extract the topics of tweets that are tagged with that emotion or morality category. 
We apply BERTopic to tweets within the three-day time window before and after the change, as discussions quickly die on social media~\cite{leskovec2009meme}. For example, for the Black Lives Matter (BLM) protests starting on 2020-05-26, we extract the topics from tweets posted between 05-23 to 05-25 to develop a baseline and then separately extract the topics between 05-26 to 05-28. By comparing the top 10 baseline topics before the change point with those after the change point, we determine the new topics that emerged after the change points that are possibly relevant to the event. See Table~\ref{tab:topic_identification} in Appendices for examples.

For preprocessing, we remove URLs and name mentions, transform emojis to their textual descriptions, and split hashtags into individual words. We use the Sentence-BERT ``all-MiniLM-L6-v2'' model \cite{reimers-2019-sentence-bert} to directly embed the processed English tweets, and ``sentence-camembert-large'' \cite{martin2020camembert} for French tweets. After topic modeling, we remove stopwords in the learned topic keywords. 
With each emerging topic, we manually verify if there is an associated offline event by examining the tweets belonging to this topic and by searching related news articles. Such manual verification is a necessary and common practice event detection literature \cite{morabia-etal-2019-sedtwik}.

\subsection{Evaluation of the Proposed Method} \label{section_eval}
Similar to prior work~\cite{10.1145/2396761.2396785,morabia-etal-2019-sedtwik}, we use precision and duplicated event rate (DERate) to evaluate our method. Recall is not used because we cannot annotate every tweet to obtain an exhaustive list of events. Precision is the fraction of detected events that are related to realistic events~\cite{Allan1998}. We manually verify each event by searching news with topic keywords associated with each change point, which is common practice in event detection research. A false positive is a change point cannot be explained by any topic and/or be related to any real event. 
Another metric is the Duplicate Event Rate (DERate), the percentage of duplicate detected events among all realistic events detected~\cite{10.1145/2396761.2396785}. We define it as the fraction of emotion and morality categories (out of 21)  that detected the same event. The higher DERate shows better confidence. Our precision and DERate for the three datasets are shown in Table~\ref{tab:pipeline_evaluation}, and \textit{are comparable to prior  works}~\cite{morabia-etal-2019-sedtwik}.

\begin{table}[h]
    \centering
    \caption{Evaluation of the Proposed Method}
    \resizebox{0.7\linewidth}{!}{
    \begin{tabular}{c|c|c|c|c}
    \Xhline{1pt}
        \textbf{Dataset} & \textbf{Precision} & \textbf{DERate} & \textbf{\# Change Points} & \textbf{\begin{tabular}[c]{@{}c@{}}Change Point\\ Confidence\end{tabular}} \\
        \hline
        2020 LA Data & 0.84 & 0.18 & 54 & 0.94 $\pm$ 0.12 \\
        2022 French Election Data & 0.83 & 0.16 & 40 & 0.85 $\pm$ 0.19 \\
        2022 Abortion Data & 0.82 & 0.16 & 99 & 0.96 $\pm$ 0.12 \\
    \Xhline{1pt}
    \end{tabular}}
    \label{tab:pipeline_evaluation}
\end{table}


\section{Disentangling Socio-political Events in 2020 Los Angeles}


\begin{figure}[!htbp]
\centering
\includegraphics[width=0.99\linewidth]{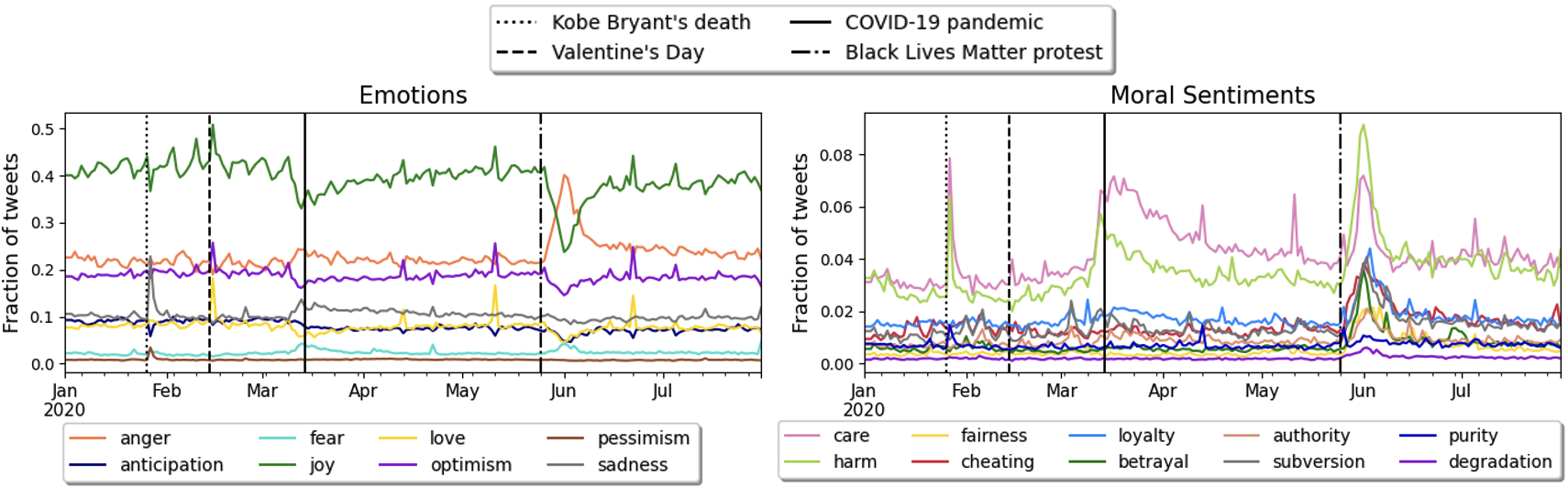} 
\caption{Time series of emotions and moral sentiments in Los Angeles Tweets from January 1 to August 1, 2020. We show the daily fraction of tweets with different affect labels. The notable peaks and dips in the time series can be associated with the external events marked as vertical lines.}
\label{fig_ts}
\end{figure}

The year of 2020 was a particularly challenging period in the city of Los Angeles. In addition to the world-wide pandemic, which 
led to a national lockdown mid-March, political primaries were also taking place during this time period, which also saw one of the largest social justice protests triggered by the murder of George Floyd in police custody, as well as the death of a beloved sports icon. These developments had a profound impact, as demonstrated by the many rises and dips in emotions and moral sentiments. Time series of the aggregate affect from January to August 2020 (Fig.~\ref{fig_ts}) shows complex dynamics with seasonal variation (weekly cycles in \textit{joy}), short-term bursts (spike in \textit{love} on Valentine's Day), and long-term changes in emotions and moral sentiments. 

\begin{figure}[h]
\centering
\includegraphics[width=1\linewidth]{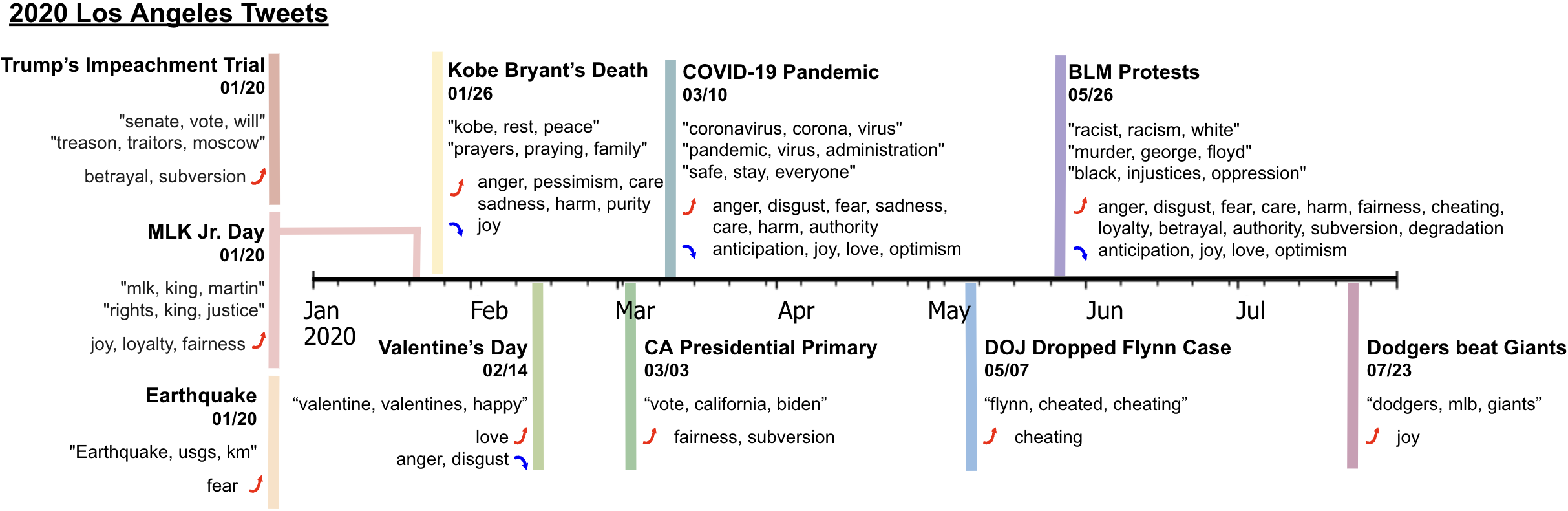} 
\caption{Top events, their associated topics and corresponding emotional and moral reactions detected in 2020 Los Angeles data. See the full list of events in Table~\ref{tab:la_reactions} in Appendices.
}
\label{fig:timeline_LA}
\end{figure}

We ran the proposed pipeline to detect and explain the online emotional reactions to events (see Table~\ref{tab:la_reactions} in Appendices for a full list of detected events). Figure~\ref{fig:timeline_LA} shows that our method is able to identify key events such as the COVID-19 pandemic and the BLM protests. We see the complex reactions to the pandemic along the multiple dimensions of emotions and morality. The unsupervised method also enables us to discover reactions to smaller events that might be easily missed, such as earthquakes and baseball playoffs. 
We also show that running BERTopic on tweets posted near the event reveals the relevant discussion topics.
Further, because we detect changes separately in each emotion, we can disentangle events based on different emotional reactions, even when they take place on the same day: Trump's impeachment trial was associated with an increase in \textit{betrayal} and \textit{subversion}, MLK Day with \textit{joy}, \textit{loyalty} and \textit{fairness}, and an earthquake with \textit{fear}.

\begin{figure}[h]
\centering
\includegraphics[width=1\linewidth]{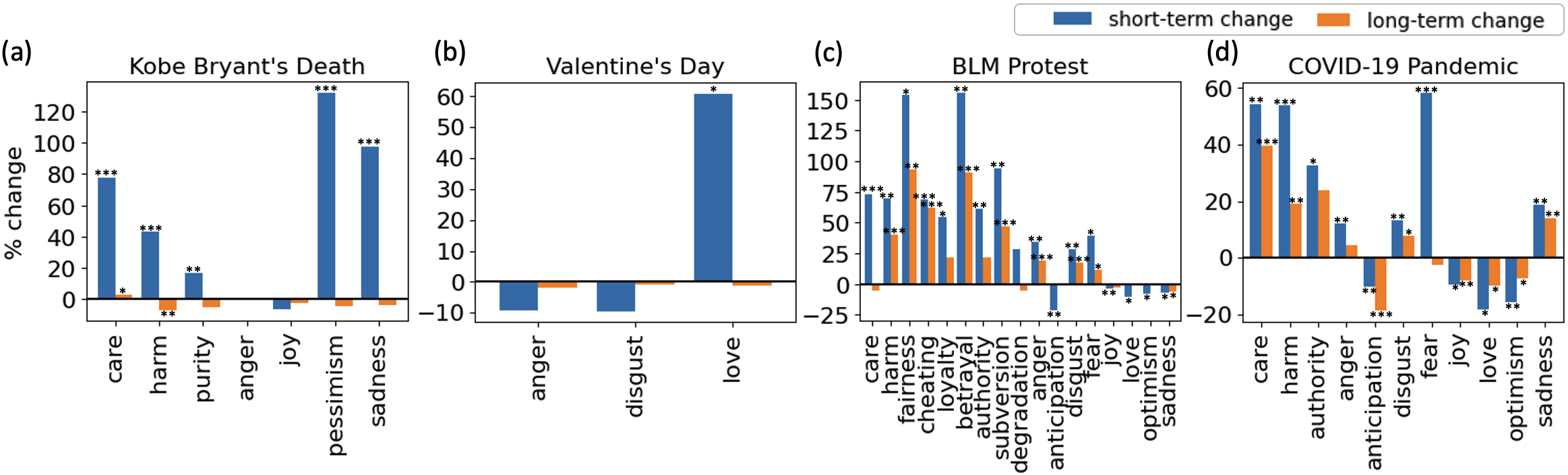} 
\caption{Short-term and long-term changes of emotions and moral sentiments around four events in 2020 Los Angeles data. Asterisks indicate significance values: * (p-value < 0.05), ** (p-value < 0.01), *** (p-value < 0.001), and no asterisk indicates p-value $\geq$ 0.05.} 
\label{fig:analysis_la}
\end{figure}

Our proposed method enables us to study collective reactions to events along multiple dimensions of affect. For example, the BLM protests were associated with 16 different emotional and moral changes. We quantify the short-term and long-term percent change in the corresponding collective affect before and after the event for four of the most impactful events (Fig.~\ref{fig:analysis_la}). 
Consistent with our intuition, Kobe Bryant's Death was associated with a short-term increase in \textit{pessimism} and \textit{sadness} and a decrease in \textit{joy}, as well as a short-term rise in moral language related to \textit{care} and \textit{harm}. In contrast, Valentine's Day brought a short-term increase in \textit{love} and a decrease in \textit{anger} and \textit{disgust}. No long-term changes were seen with these events. On the other hand, the BLM protests was associated with complex short- and long-term changes in affect. We observe increases in negative emotions and decreases in positive emotions. In addition, compared to other three events, we see greater increases in moral sentiments. The moral concerns about \textit{fairness} and \textit{betrayal} had especially increased, expressing a deep sense of the injustice and betrayal in George Floyd's death.

The COVID-19 outbreak triggered a cascade of events aimed at mitigating the pandemic that were associated with complex short-term and long-term changes in affect (Fig.~\ref{fig:analysis_la}d). People expressed more \textit{anger}, \textit{disgust}, \textit{sadness}, and more significantly, \textit{fear}. Positive emotions like \textit{joy} and \textit{love} simultaneously decreased, both in the short-term and the long-term. People also expressed more moral sentiments like \textit{care} such as in ``Stay safe. We thank you'', as well as more \textit{harm} blaming the virus. 
Interestingly, the moral language around \textit{authority} also increased, possibly due to new policies such as lockdowns to mitigate the pandemic (e.g. ``I think governor Newsom is doing a great job...''), and some were critical of government's response,  e.g., ``we need leadership not a politician''. 
Because Twitter users are predominantly liberal, and this dataset is collected in Los Angeles (91\% users to be liberal and 9\% conservative), we found the emotional and moral reactions to these events reflecting liberal perspectives.



\begin{figure}[h]
\centering
\includegraphics[width=0.5\linewidth]{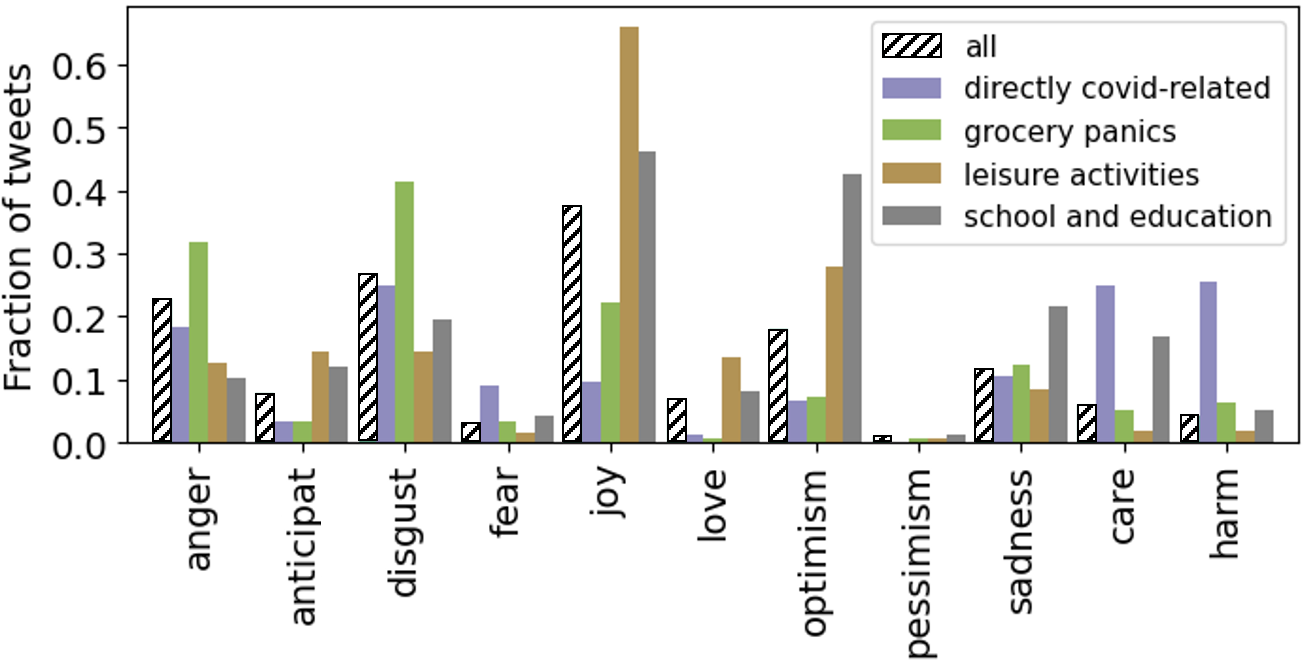} 
\caption{Emotions and moral sentiments expressed in COVID-related topics during the two weeks after WHO announcement of the pandemic on 2020-03-11. The topics are COVID (``coronavirus, corona, virus''); grocery panics (``grocery, groceries, shelves'', ``water, dasani, hydro'', and ``toilet, paper, rolls'');  leisure activities (``episode, episodes, show'', ``cook, cooking, cookout'',``tickets, ticket, selling''); and education (``teachers, students, learning'', ``schools, lausd, classes'', ``schools, lausd, closed''). 
}
\label{fig:covid_emot}
\end{figure}

Next, we take a deep dive in the COVID-19 emotion analysis, and show the benefit of disentangling emotional reactions by disaggregating topics. We select four top categories discussed and group related topics into these categories: directly covid-related topics, grocery panics, leisure activities and school and education. 
We study emotions and moral expressions aggregated in all the tweets, as well as in these topic categories (Fig. \ref{fig:covid_emot}). 
We find that aggregating emotions from all tweets can give misleading impressions. Positive emotions like \textit{joy} were mostly expressed in all tweets (aggregated), but in fact they were mostly dominated by people talking about leisure activities. In COVID-related tweets, few positive emotions were expressed. \textit{Anger} and \textit{disgust} were higher in topics about grocery panics than in topics directly related to COVID. Another example is the expression of \textit{care} and \textit{harm} moral sentiments. Their expression was diluted by other topics in aggregate tweets. After disaggregating, they were largely expressed in directly COVID-related tweets. These results suggest that during times of maximal crisis and uncertainty, people find outlets for positive emotions. They also demonstrate the importance of disaggregating by topics when studying specific issues {like COVID-19 that cover a multitude of fine-grained sub-topics}.

\section{Unfolding the Evolution of Abortion Rights Discussion}

\begin{figure}[h]
\centering
\includegraphics[width=1\linewidth]{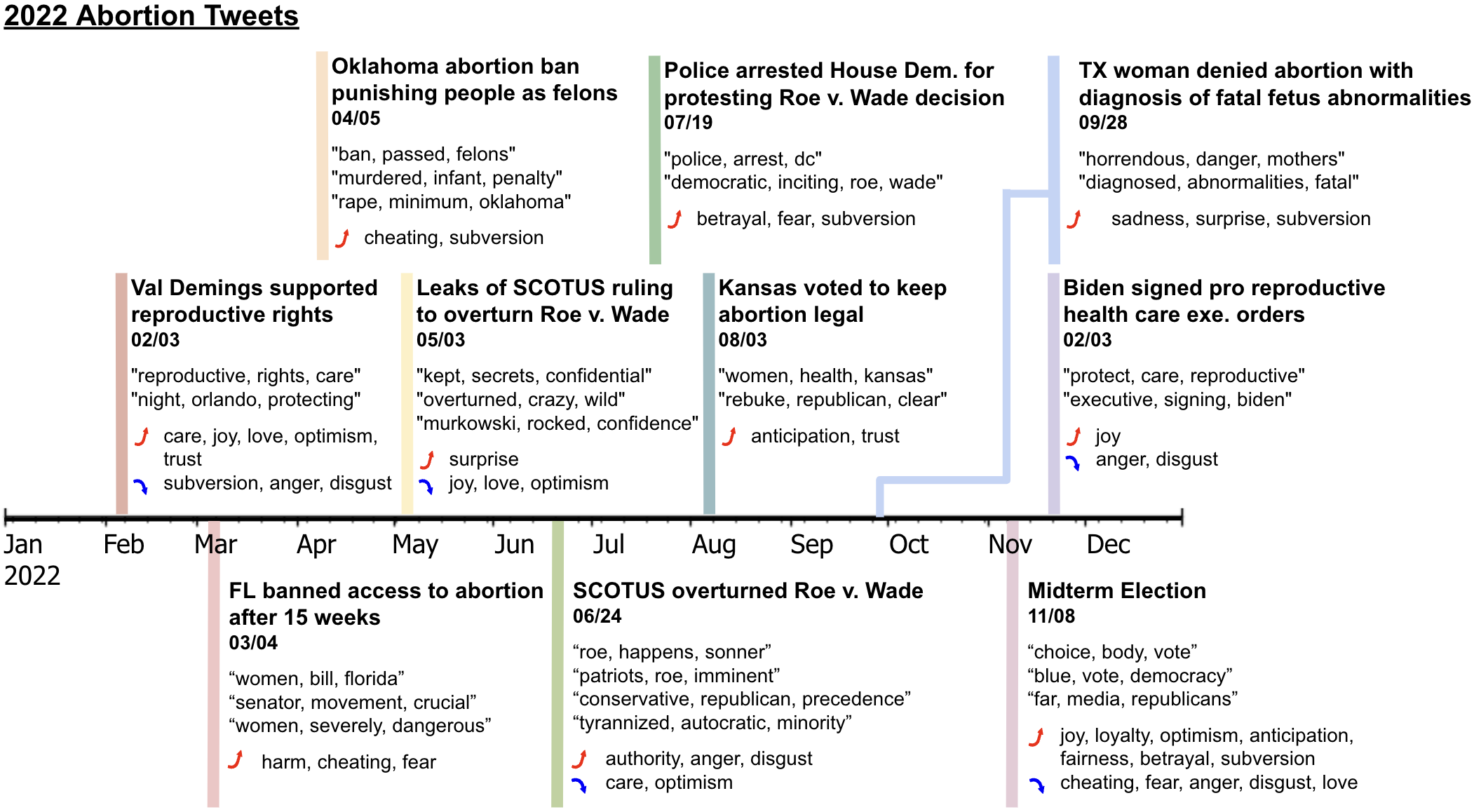} 
\caption{Top events, their associated topics and corresponding emotional and moral reactions detected in 2022 Abortion data. See the full list of events in Table~\ref{tab:abortion_reactions} in Appendices.
}
\label{fig:timeline_abortion}
\end{figure}

\begin{figure}[]
\centering
\includegraphics[width=1\linewidth]{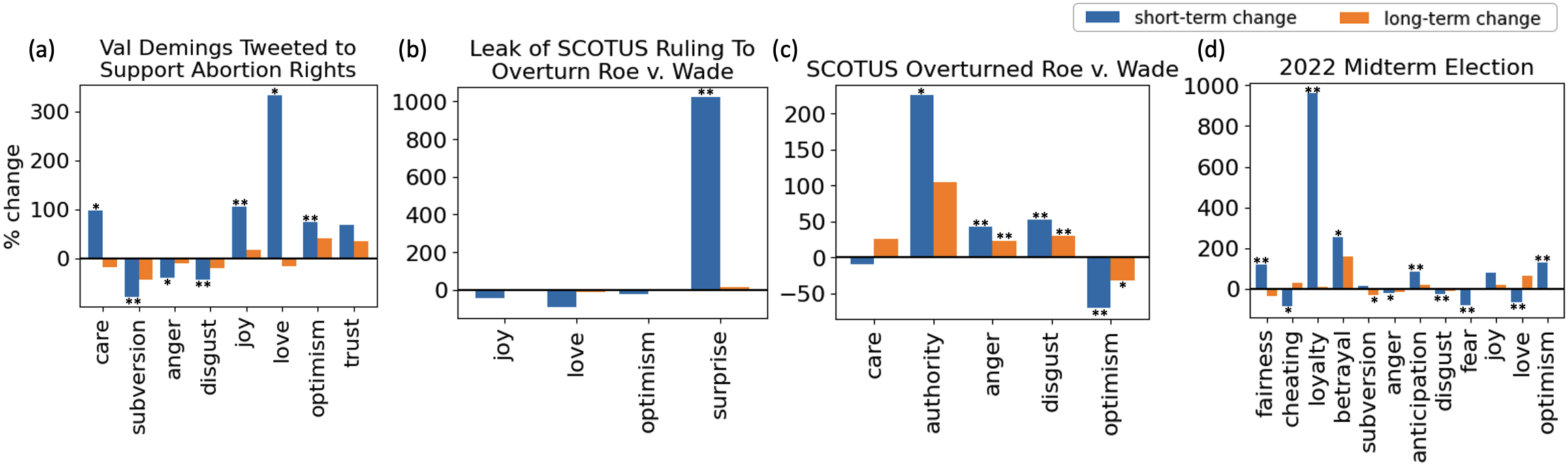} 
\caption{Short-term and long-term changes of emotions and moral sentiments of some top events in 2022 Abortion Tweets. Asterisks indicate significance values: * (p-value < 0.05), ** (p-value < 0.01), *** (p-value < 0.001), and no asterisk indicates p-value $\geq$ 0.05.
}
\label{fig:analysis_abortion}
\end{figure}

Abortion is one of the most politically charged issues in the U.S. The debate was especially intense in 2022, when the Supreme Court of the United States (SCOTUS) struck down federal protections for abortion rights in its Dobbs v. Jackson Women’s Health Organization ruling on June 24, 2022. This decision overturned the nearly 50 year old precedent set by the Roe vs Wade decision, which guaranteed women in U.S. access to abortion. In the 2022 Abortion Tweets data, we have identified a total of 24 different events, unfolding the evolution of abortion discussions on Twitter (Figure~\ref{fig:timeline_abortion}). See a full list of events detected in Table~\ref{tab:abortion_reactions}. We detected major events, such as the leak of SCOTUS ruling on May 3rd, to which people expressed over 1000\% more \textit{surprise} (Fig.~\ref{fig:analysis_abortion}b), and the overturning of Roe v. Wade on June 24th, followed by a surge of strong emotional reactions, including increasing \textit{anger} and \textit{disgust}, and a dip in \textit{care} and \textit{optimism} (Fig.~\ref{fig:analysis_abortion}c). This finding is consistent with the previous work by \citeauthor{rao2023tracking}. Along with the SCOTUS ruling, multiple states had issued abortion bans, which further provoked online debates. We detected the issuing of abortion bans in Florida and Oklahoma on March 4th and April 5th, respectively, and observed surging negative moral sentiments, such as \textit{harm}, \textit{cheating} and \textit{subversion}. In the contrast, when Kansas voted to keep abortion legal on August 3rd, \textit{anticipation} and \textit{trust} was expressed.

This historical SCOTUS ruling did not only elicit temporary emotional reactions, but its impact is long term. We detected multiple protests as well as police arrests of protesters, accompanied by online discussions expressing \textit{betrayal}, \textit{subversion}, \textit{fear} and \textit{sadness}. There were also viral stories of individuals experiencing health and legal crises because of the ruling. One example is a woman in Texas who was denied an abortion even with diagnosis of fatal fetal abnormalities. Increasing \textit{surprise}, \textit{sadness} and \textit{subversion} were detected in these discussions (Figure~\ref{fig:timeline_abortion} and Table~\ref{tab:abortion_reactions}).

Another important and long-term impact of this ruling was in the 2022 US Midterm Election. Many congressmen pushed out agendas related to abortion rights. Some Democrats such as Biden and Val Demings stated their support for abortion rights (Figure~\ref{fig:timeline_abortion}). In response, we detected increasing \textit{care}, \textit{love}, and \textit{trust} in the dataset. On the other hand, some Republicans implied a pro-life stance  (e.g. event 3 and 11 in Table~\ref{tab:abortion_reactions}). When the Midterm Election happened on November 8th, we observed a complex emotional and moral reaction (Fig.~\ref{fig:analysis_abortion}d). In general, positive emotions and moral sentiments such as \textit{loyalty}, \textit{anticipation} and \textit{optimism} increased, whereas negative emotions decreased, indicating people were hoping for their favored election results. 

The 24 events we detected in this dataset include major and momentous events, small events such as people re-sharing a viral tweet, as well as events that took place closely in time and even on the same day (Table~\ref{tab:abortion_reactions}). This demonstrates again the effectiveness of our proposed method to automatically detect emotional reactions. Last but not least, the emotional and moral reactions to most of these events are left-leaning, similar to that in the Los Angeles dataset. This can be explained by that Twitter users are predominantly liberal. In this dataset, we found 72\% users to be liberal and 28\% conservative \cite{rao2023tracking}.

\section{Understanding Emotion Dynamics in the 2022 French Election}
\begin{figure}[h]
\centering
\includegraphics[width=0.9\linewidth]{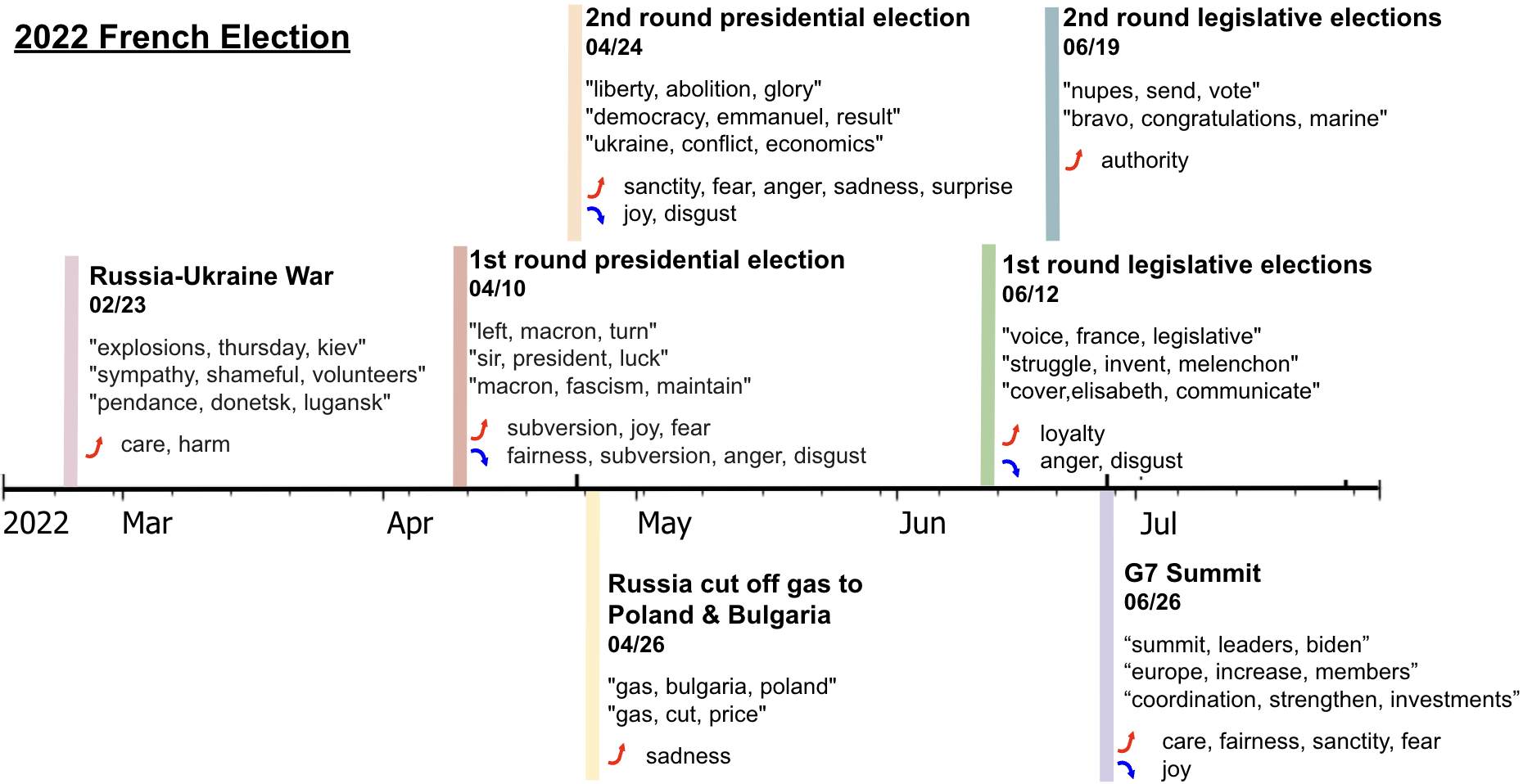} 
\caption{Top events, their associated topics and corresponding emotional and moral reactions detected in 2022 French Election data. See the full list of events in Table~\ref{tab:french_reactions} in Appendices
}
\label{fig:timeline_french}
\end{figure}

Our third case study is the 2022 French Election, mainly among Emmanuel Macron, Marine Le Pen and Jean-Luc Mélenchon. This election took place close in time to the Russia-Ukraine War, which started in February 2022, and the G7 Summit in June, adding complications to the study of emotional reactions in the online population. Previous reports have shown that the Russia-Ukraine war had significant implications for the election campaigns, especially a positive effect on Macron's polling \cite{Schofield_2022}. On the Twitter dataset, we are able to detect both the first and second rounds of presidential elections on April 10th and April 24th, as well as the two rounds of legislative elections on June 12th and June 19th. In addition, we have also detected the G7 summit and various events related to the Russia-Ukraine war, including the initial invasion, the sinking of a Russian warship, and Russia cutting off natural gas to East Europe (Figure~\ref{fig:timeline_french} and Table~\ref{tab:french_reactions}). The online discussions about the election also involved topics about the Russia-Ukraine war, such as ``ukraine, conflict, economics''.

\begin{figure}[h]
\centering
\includegraphics[width=1\linewidth]{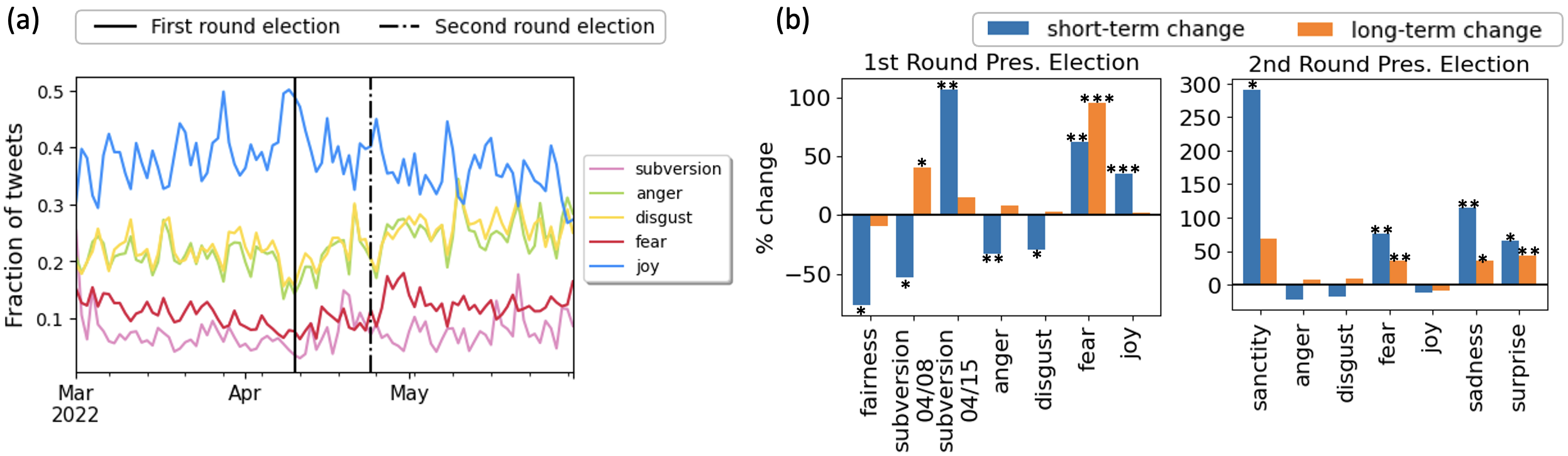} 
\caption{(a) Time series of emotions and moral sentiments with significant changes during the 2022 French Election. (b) Short-term and long-term changes of emotions and moral sentiments around first and second rounds of 2022 French Election. Asterisks indicate significance values: * (p-value < 0.05), ** (p-value < 0.01), *** (p-value < 0.001), and no asterisk indicates p-value $\geq$ 0.05.
}
\label{fig:analysis_french}
\end{figure}

Next, we analyze the dynamics of emotions and moral sentiments during the presidential election cycle. The change points in different emotion and moral categories are detected in a relatively wide range of dates, from 04/07 to 04/15 for the first round, which actually happened on 04/10, and from 04/21 to 04/26 for the second round, which took place on 04/24. This indicates the convoluted influence of the voting rounds. We found an interesting pattern, that \textit{positive emotions and moral sentiments increased before each voting round, but negative sentiments surged up right after the voting round}. This is reflected in the time series with the zigzag shapes in Figure~\ref{fig:analysis_french}a. For example, \textit{anger} and \textit{subversion} first surged then dipped between two rounds, whereas \textit{joy} moved in the opposite way, dipping first and then surging between the two rounds. This indicates that people were showing hope and support before each voting round, but started to reflect and criticize after the voting finished and results came out.
Fig.~\ref{fig:analysis_french}b also reveals similar patterns. The most interesting example is the changes in subversion. There were two change points detected in subversion. It decreased by 52.77\% on 04/08 before the first voting round, but significantly increased by 107.12\% on 04/15 after, and then dropped again right before the second round election started. These change points are consistent with the time series pattern shown in Fig.~\ref{fig:analysis_french}a. In addition, we also observe the significant increase in fear after each voting round. These again show people's hope before voting, and negativity and fear after voting results revealed. 

In the French Election dataset, we have successfully detected all events related to the election and those related to the war. Furthermore, we are able to disentangle the complicated dynamics of emotion and moral changes, benefiting from the proposed change point detection method.

\section{Conclusion}
In this work, we have demonstrated the effectiveness of an unsupervised method to detect and measure public reactions to newsworthy events. We applied our method to three large Twitter datasets. We have disentangled the dynamics of online emotions during a time period punctuated by complex social, health, and political events in the 2020 Los Angeles data, studied the evolution of abortion rights discussion along with the overturning of Roe v. Wade in 2022, and revealed the interesting dynamics of emotions during the 2022 French election. We showed that our method can discover major and subtle events and even events that happened closely in time, and measure emotional and moral reactions to these events. In addition, we demonstrated, using the example of COVID-19 outbreak, the importance of disaggregating by topics when further understand the complex impact of some major events. 
Together, these results suggests the potential of using social media data for sensing and tracking of public reactions to events, as well as discovering significant events that may have been missed by traditional news sources.

\paragraph{Limitations and Future Works} 
First, Twitter users are predominantly liberal. Hence the emotional reaction analysis we present inevitably include more left-leaning perspectives, especially in the U.S. tweet datasets. In the future we plan to split data by demographic groups and analyze the heterogeneity in different user groups. Second, with our emotional reaction detection method, when there is a change point that is a dip, we cannot use topic modeling to explain it, as a dip in the emotion or moral sentiment indicates a decrease of discussion related to an event. However, usually the decrease of some emotions is accompanied by the increase of some other emotions, and we can study the tweets tagged with the surged emotions to understand the topics.

In future works we plan to move forward to causal analysis. In principle, the offline events \textit{cause} the online emotional and moral reactions. However, there are many confounding factors in this causal process. For example, the same event might pose very different effects on heterogeneous online populations. We plan to expand this work by performing causal analysis to further disentangle the causal relationships between offline events and online emotions, and to measure the heterogeneous effects on different online populations.


\begin{acks}
This work is supported in part by AFOSR under grants FA9550-22-1-0380 \& FA9550-20-1-0224, and DARPA under contract HR001121C0168. The authors thank Eugene Jang and Yuanfeixue Nan from University of Southern California for helping with emotion and moral foundation annotations.

\end{acks}

\bibliographystyle{ACM-Reference-Format}
\bibliography{ref}


\begin{thebibliography}{47}


\ifx \showCODEN    \undefined \def \showCODEN     #1{\unskip}     \fi
\ifx \showDOI      \undefined \def \showDOI       #1{#1}\fi
\ifx \showISBNx    \undefined \def \showISBNx     #1{\unskip}     \fi
\ifx \showISBNxiii \undefined \def \showISBNxiii  #1{\unskip}     \fi
\ifx \showISSN     \undefined \def \showISSN      #1{\unskip}     \fi
\ifx \showLCCN     \undefined \def \showLCCN      #1{\unskip}     \fi
\ifx \shownote     \undefined \def \shownote      #1{#1}          \fi
\ifx \showarticletitle \undefined \def \showarticletitle #1{#1}   \fi
\ifx \showURL      \undefined \def \showURL       {\relax}        \fi
\providecommand\bibfield[2]{#2}
\providecommand\bibinfo[2]{#2}
\providecommand\natexlab[1]{#1}
\providecommand\showeprint[2][]{arXiv:#2}

\bibitem[Adams et~al\mbox{.}(2007)]%
        {adams2007bayesian}
\bibfield{author}{\bibinfo{person}{Ryan~Prescott Adams} {et~al\mbox{.}}} \bibinfo{year}{2007}\natexlab{}.
\newblock \showarticletitle{Bayesian online changepoint detection}.
\newblock \bibinfo{journal}{\emph{arXiv preprint arXiv:0710.3742}} (\bibinfo{year}{2007}).
\newblock


\bibitem[Aldous et~al\mbox{.}(2022)]%
        {aldous2022measuring}
\bibfield{author}{\bibinfo{person}{Kholoud~Khalil Aldous} {et~al\mbox{.}}} \bibinfo{year}{2022}\natexlab{}.
\newblock \showarticletitle{Measuring 9 emotions of news posts from 8 news organizations across 4 social media platforms for 8 months}.
\newblock \bibinfo{journal}{\emph{ACM Transactions on Social Computing (TSC)}} \bibinfo{volume}{4}, \bibinfo{number}{4} (\bibinfo{year}{2022}), \bibinfo{pages}{1--31}.
\newblock


\bibitem[Alhuzali et~al\mbox{.}(2021)]%
        {alhuzali-ananiadou-2021-spanemo}
\bibfield{author}{\bibinfo{person}{Hassan Alhuzali} {et~al\mbox{.}}} \bibinfo{year}{2021}\natexlab{}.
\newblock \showarticletitle{{S}pan{E}mo: Casting Multi-label Emotion Classification as Span-prediction}. In \bibinfo{booktitle}{\emph{ECACL}}. \bibinfo{publisher}{ACL}, \bibinfo{pages}{1573--1584}.
\newblock


\bibitem[Allan et~al\mbox{.}(1998)]%
        {Allan1998}
\bibfield{author}{\bibinfo{person}{James Allan} {et~al\mbox{.}}} \bibinfo{year}{1998}\natexlab{}.
\newblock \showarticletitle{{Topic Detection and Tracking Pilot Study Final Report}}.
\newblock  (\bibinfo{year}{1998}).
\newblock
\urldef\tempurl%
\url{https://doi.org/10.1184/R1/6626252.v1}
\showDOI{\tempurl}


\bibitem[Babac(2022)]%
        {babac2022emotion}
\bibfield{author}{\bibinfo{person}{Marina~Bagi{\'c} Babac}.} \bibinfo{year}{2022}\natexlab{}.
\newblock \showarticletitle{Emotion analysis of user reactions to online news}.
\newblock \bibinfo{journal}{\emph{Information Discovery and Delivery}} \bibinfo{number}{ahead-of-print} (\bibinfo{year}{2022}).
\newblock


\bibitem[Barbieri et~al\mbox{.}(2022)]%
        {xlm2021barbieri}
\bibfield{author}{\bibinfo{person}{Francesco Barbieri}, \bibinfo{person}{Luis~Espinosa Anke}, {and} \bibinfo{person}{Jose Camacho-Collados}.} \bibinfo{year}{2022}\natexlab{}.
\newblock \showarticletitle{{XLM-T}: Multilingual language models in twitter for sentiment analysis and beyond}. In \bibinfo{booktitle}{\emph{LREC}}. \bibinfo{pages}{258--266}.
\newblock


\bibitem[Blei et~al\mbox{.}(2003)]%
        {blei2003latent}
\bibfield{author}{\bibinfo{person}{David~M Blei} {et~al\mbox{.}}} \bibinfo{year}{2003}\natexlab{}.
\newblock \showarticletitle{Latent dirichlet allocation}.
\newblock \bibinfo{journal}{\emph{JMLR}} \bibinfo{volume}{3}, \bibinfo{number}{Jan} (\bibinfo{year}{2003}), \bibinfo{pages}{993--1022}.
\newblock


\bibitem[Bollen et~al\mbox{.}(2011)]%
        {bollen2011twitter}
\bibfield{author}{\bibinfo{person}{Johan Bollen} {et~al\mbox{.}}} \bibinfo{year}{2011}\natexlab{}.
\newblock \showarticletitle{Twitter mood predicts the stock market}.
\newblock \bibinfo{journal}{\emph{Journal of computational science}} \bibinfo{volume}{2}, \bibinfo{number}{1} (\bibinfo{year}{2011}), \bibinfo{pages}{1--8}.
\newblock


\bibitem[Cao et~al\mbox{.}(2021)]%
        {cao2021knowledge}
\bibfield{author}{\bibinfo{person}{Yuwei Cao} {et~al\mbox{.}}} \bibinfo{year}{2021}\natexlab{}.
\newblock \showarticletitle{Knowledge-preserving incremental social event detection via heterogeneous gnns}. In \bibinfo{booktitle}{\emph{In WWW}}. \bibinfo{pages}{3383--3395}.
\newblock


\bibitem[Chang et~al\mbox{.}(2023)]%
        {chang2023roeoverturned}
\bibfield{author}{\bibinfo{person}{Rong-Ching Chang}, \bibinfo{person}{Ashwin Rao}, \bibinfo{person}{Qiankun Zhong}, \bibinfo{person}{Magdalena Wojcieszak}, {and} \bibinfo{person}{Kristina Lerman}.} \bibinfo{year}{2023}\natexlab{}.
\newblock \showarticletitle{\# RoeOverturned: Twitter Dataset on the Abortion Rights Controversy}. In \bibinfo{booktitle}{\emph{Proceedings of the International AAAI Conference on Web and Social Media}}, Vol.~\bibinfo{volume}{17}. \bibinfo{pages}{997--1005}.
\newblock


\bibitem[Cody et~al\mbox{.}(2015)]%
        {cody2015climate}
\bibfield{author}{\bibinfo{person}{Emily~M Cody} {et~al\mbox{.}}} \bibinfo{year}{2015}\natexlab{}.
\newblock \showarticletitle{Climate change sentiment on Twitter: An unsolicited public opinion poll}.
\newblock \bibinfo{journal}{\emph{PloS one}} \bibinfo{volume}{10}, \bibinfo{number}{8} (\bibinfo{year}{2015}), \bibinfo{pages}{e0136092}.
\newblock


\bibitem[Dieng et~al\mbox{.}(2020)]%
        {dieng-etal-2020-topic}
\bibfield{author}{\bibinfo{person}{Adji~B. Dieng} {et~al\mbox{.}}} \bibinfo{year}{2020}\natexlab{}.
\newblock \showarticletitle{Topic Modeling in Embedding Spaces}.
\newblock \bibinfo{journal}{\emph{TACL}}  \bibinfo{volume}{8} (\bibinfo{year}{2020}), \bibinfo{pages}{439--453}.
\newblock
\urldef\tempurl%
\url{https://doi.org/10.1162/tacl_a_00325}
\showDOI{\tempurl}


\bibitem[Dodds et~al\mbox{.}(2022)]%
        {dodds2022fame}
\bibfield{author}{\bibinfo{person}{Dodds} {et~al\mbox{.}}} \bibinfo{year}{2022}\natexlab{}.
\newblock \showarticletitle{Fame and Ultrafame: Measuring and comparing daily levels of ‘being talked about’ for United States’ presidents, their rivals, God, countries, and K-pop.}
\newblock \bibinfo{journal}{\emph{JQD:DM}}  \bibinfo{volume}{2} (\bibinfo{date}{Feb.} \bibinfo{year}{2022}).
\newblock
\urldef\tempurl%
\url{https://doi.org/10.51685/jqd.2022.004}
\showDOI{\tempurl}


\bibitem[Dodds et~al\mbox{.}(2011)]%
        {dodds2011temporal}
\bibfield{author}{\bibinfo{person}{Peter Dodds} {et~al\mbox{.}}} \bibinfo{year}{2011}\natexlab{}.
\newblock \showarticletitle{Temporal patterns of happiness and information in a global social network: Hedonometrics and Twitter}.
\newblock \bibinfo{journal}{\emph{PloS one}} \bibinfo{volume}{6}, \bibinfo{number}{12} (\bibinfo{year}{2011}), \bibinfo{pages}{e26752}.
\newblock


\bibitem[Garten et~al\mbox{.}(2018)]%
        {8b2871f503a14011ae81e6ab1664a638}
\bibfield{author}{\bibinfo{person}{Justin Garten} {et~al\mbox{.}}} \bibinfo{year}{2018}\natexlab{}.
\newblock \showarticletitle{Dictionaries and distributions: Combining expert knowledge and large scale textual data content analysis: Distributed dictionary representation}.
\newblock \bibinfo{journal}{\emph{Behavior Research Methods, Instruments, and Computers}} \bibinfo{volume}{50}, \bibinfo{number}{1} (\bibinfo{date}{1 Feb.} \bibinfo{year}{2018}), \bibinfo{pages}{344--361}.
\newblock
\showISSN{1554-351X}
\urldef\tempurl%
\url{https://doi.org/10.3758/s13428-017-0875-9}
\showDOI{\tempurl}


\bibitem[Golder et~al\mbox{.}(2011)]%
        {golder2011diurnal}
\bibfield{author}{\bibinfo{person}{Scott~A Golder} {et~al\mbox{.}}} \bibinfo{year}{2011}\natexlab{}.
\newblock \showarticletitle{Diurnal and seasonal mood vary with work, sleep, and daylength across diverse cultures}.
\newblock \bibinfo{journal}{\emph{Science}} \bibinfo{volume}{333}, \bibinfo{number}{6051} (\bibinfo{year}{2011}), \bibinfo{pages}{1878--1881}.
\newblock


\bibitem[Grootendorst(2022)]%
        {grootendorst2022bertopic}
\bibfield{author}{\bibinfo{person}{Maarten Grootendorst}.} \bibinfo{year}{2022}\natexlab{}.
\newblock \showarticletitle{BERTopic: Neural topic modeling with a class-based TF-IDF procedure}.
\newblock \bibinfo{journal}{\emph{arXiv preprint arXiv:2203.05794}} (\bibinfo{year}{2022}).
\newblock


\bibitem[Guo et~al\mbox{.}(2023a)]%
        {guo2023data}
\bibfield{author}{\bibinfo{person}{Siyi Guo} {et~al\mbox{.}}} \bibinfo{year}{2023}\natexlab{a}.
\newblock \showarticletitle{A Data Fusion Framework for Multi-Domain Morality Learning}. In \bibinfo{booktitle}{\emph{In ICWSM-2023}}, Vol.~\bibinfo{volume}{17}. \bibinfo{pages}{281--291}.
\newblock


\bibitem[Guo et~al\mbox{.}(2023b)]%
        {guo2023measure}
\bibfield{author}{\bibinfo{person}{Siyi Guo} {et~al\mbox{.}}} \bibinfo{year}{2023}\natexlab{b}.
\newblock \showarticletitle{Measure Online Emotional Reactions to Events}. In \bibinfo{booktitle}{\emph{Proceedings of ASONAM '23}}. \bibinfo{publisher}{Association for Computing Machinery}, \bibinfo{address}{New York, NY, USA}.
\newblock


\bibitem[Haidt et~al\mbox{.}(2007)]%
        {haidt2007moral}
\bibfield{author}{\bibinfo{person}{Jonathan Haidt} {et~al\mbox{.}}} \bibinfo{year}{2007}\natexlab{}.
\newblock \showarticletitle{The moral mind: How five sets of innate intuitions guide the development of many culture-specific virtues, and perhaps even modules}.
\newblock \bibinfo{journal}{\emph{The innate mind}}  \bibinfo{volume}{3} (\bibinfo{year}{2007}), \bibinfo{pages}{367--391}.
\newblock


\bibitem[Hauthal et~al\mbox{.}(2019)]%
        {hauthal2019analyzing}
\bibfield{author}{\bibinfo{person}{Eva Hauthal} {et~al\mbox{.}}} \bibinfo{year}{2019}\natexlab{}.
\newblock \showarticletitle{Analyzing and visualizing emotional reactions expressed by emojis in location-based social media}.
\newblock \bibinfo{journal}{\emph{ISPRS International Journal of Geo-Information}} \bibinfo{volume}{8}, \bibinfo{number}{3} (\bibinfo{year}{2019}), \bibinfo{pages}{113}.
\newblock


\bibitem[He et~al\mbox{.}(2022)]%
        {he2022infusing}
\bibfield{author}{\bibinfo{person}{Zihao He}, \bibinfo{person}{Negar Mokhberian}, {and} \bibinfo{person}{Kristina Lerman}.} \bibinfo{year}{2022}\natexlab{}.
\newblock \showarticletitle{Infusing Knowledge from Wikipedia to Enhance Stance Detection}. In \bibinfo{booktitle}{\emph{Proceedings of the 12th Workshop on Computational Approaches to Subjectivity, Sentiment \& Social Media Analysis}}. \bibinfo{pages}{71--77}.
\newblock


\bibitem[Hinkley(1971)]%
        {hinkley1971inference}
\bibfield{author}{\bibinfo{person}{David~V Hinkley}.} \bibinfo{year}{1971}\natexlab{}.
\newblock \showarticletitle{Inference about the change-point from cumulative sum tests}.
\newblock \bibinfo{journal}{\emph{Biometrika}} \bibinfo{volume}{58}, \bibinfo{number}{3} (\bibinfo{year}{1971}), \bibinfo{pages}{509--523}.
\newblock


\bibitem[Hoover et~al\mbox{.}(2020)]%
        {Hoover2020moral}
\bibfield{author}{\bibinfo{person}{Joe Hoover} {et~al\mbox{.}}} \bibinfo{year}{2020}\natexlab{}.
\newblock \showarticletitle{Moral Foundations Twitter Corpus: A Collection of 35k Tweets Annotated for Moral Sentiment}.
\newblock \bibinfo{journal}{\emph{Social Psychological and Personality Science}} \bibinfo{volume}{11}, \bibinfo{number}{8} (\bibinfo{year}{2020}), \bibinfo{pages}{1057--1071}.
\newblock
\urldef\tempurl%
\url{https://doi.org/10.1177/1948550619876629}
\showDOI{\tempurl}


\bibitem[Jaidka et~al\mbox{.}(2020)]%
        {jaidka2020estimating}
\bibfield{author}{\bibinfo{person}{Kokil Jaidka} {et~al\mbox{.}}} \bibinfo{year}{2020}\natexlab{}.
\newblock \showarticletitle{Estimating geographic subjective well-being from Twitter: A comparison of dictionary and data-driven language methods}.
\newblock \bibinfo{journal}{\emph{PNAS}} \bibinfo{volume}{117}, \bibinfo{number}{19} (\bibinfo{year}{2020}), \bibinfo{pages}{10165--10171}.
\newblock


\bibitem[Klašnja et~al\mbox{.}(2018)]%
        {Barbera2015measuring}
\bibfield{author}{\bibinfo{person}{Marko Klašnja} {et~al\mbox{.}}} \bibinfo{year}{2018}\natexlab{}.
\newblock \showarticletitle{{Measuring Public Opinion with Social Media Data}}.
\newblock \bibinfo{publisher}{Oxford University Press}.
\newblock
\showISBNx{9780190213299}


\bibitem[Leskovec et~al\mbox{.}(2009)]%
        {leskovec2009meme}
\bibfield{author}{\bibinfo{person}{Jure Leskovec} {et~al\mbox{.}}} \bibinfo{year}{2009}\natexlab{}.
\newblock \showarticletitle{Meme-tracking and the dynamics of the news cycle}. In \bibinfo{booktitle}{\emph{Proceedings of the 15th ACM SIGKDD}}. \bibinfo{pages}{497--506}.
\newblock


\bibitem[Li et~al\mbox{.}(2012)]%
        {10.1145/2396761.2396785}
\bibfield{author}{\bibinfo{person}{Chenliang Li} {et~al\mbox{.}}} \bibinfo{year}{2012}\natexlab{}.
\newblock \showarticletitle{Twevent: Segment-Based Event Detection from Tweets} \emph{(\bibinfo{series}{CIKM '12})}. \bibinfo{publisher}{ACM}, \bibinfo{address}{New York, NY, USA}, \bibinfo{pages}{155–164}.
\newblock
\showISBNx{9781450311564}
\urldef\tempurl%
\url{https://doi.org/10.1145/2396761.2396785}
\showDOI{\tempurl}


\bibitem[Malik et~al\mbox{.}(2022)]%
        {malik2022performance}
\bibfield{author}{\bibinfo{person}{Muzamil Malik} {et~al\mbox{.}}} \bibinfo{year}{2022}\natexlab{}.
\newblock \showarticletitle{A Performance Comparison of Unsupervised Techniques for Event Detection from Oscar Tweets}.
\newblock \bibinfo{journal}{\emph{Computational Intelligence and Neuroscience}}  \bibinfo{volume}{2022} (\bibinfo{year}{2022}).
\newblock


\bibitem[Martin et~al\mbox{.}(2020)]%
        {martin2020camembert}
\bibfield{author}{\bibinfo{person}{Louis Martin}, \bibinfo{person}{Benjamin Muller}, \bibinfo{person}{Pedro Javier~Ortiz Su{\'a}rez}, \bibinfo{person}{Yoann Dupont}, \bibinfo{person}{Laurent Romary}, \bibinfo{person}{{\'E}ric~Villemonte de~la Clergerie}, \bibinfo{person}{Djam{\'e} Seddah}, {and} \bibinfo{person}{Beno{\^\i}t Sagot}.} \bibinfo{year}{2020}\natexlab{}.
\newblock \showarticletitle{CamemBERT: a Tasty French Language Mode}.
\newblock \bibinfo{journal}{\emph{Proceedings of the 58th Annual Meeting of the Association for Computational Linguistics}} (\bibinfo{year}{2020}).
\newblock


\bibitem[Mirlohi~Falavarjani et~al\mbox{.}(2021)]%
        {mirlohi2021causal}
\bibfield{author}{\bibinfo{person}{Seyed~Amin Mirlohi~Falavarjani}, \bibinfo{person}{Jelena Jovanovic}, \bibinfo{person}{Hossein Fani}, \bibinfo{person}{Ali~A Ghorbani}, \bibinfo{person}{Zeinab Noorian}, {and} \bibinfo{person}{Ebrahim Bagheri}.} \bibinfo{year}{2021}\natexlab{}.
\newblock \showarticletitle{On the causal relation between real world activities and emotional expressions of social media users}.
\newblock \bibinfo{journal}{\emph{Journal of the Association for Information Science and Technology}} \bibinfo{volume}{72}, \bibinfo{number}{6} (\bibinfo{year}{2021}), \bibinfo{pages}{723--743}.
\newblock


\bibitem[Mitchell et~al\mbox{.}(2013)]%
        {mitchell2013geography}
\bibfield{author}{\bibinfo{person}{Lewis Mitchell} {et~al\mbox{.}}} \bibinfo{year}{2013}\natexlab{}.
\newblock \showarticletitle{The geography of happiness: Connecting twitter sentiment and expression, demographics, and objective characteristics of place}.
\newblock \bibinfo{journal}{\emph{PloS one}} \bibinfo{volume}{8}, \bibinfo{number}{5} (\bibinfo{year}{2013}), \bibinfo{pages}{e64417}.
\newblock


\bibitem[Mohammad et~al\mbox{.}(2018)]%
        {mohammad-etal-2018-semeval}
\bibfield{author}{\bibinfo{person}{Saif Mohammad} {et~al\mbox{.}}} \bibinfo{year}{2018}\natexlab{}.
\newblock \showarticletitle{{S}em{E}val-2018 Task 1: Affect in Tweets}. In \bibinfo{booktitle}{\emph{Proc. 12th Int. Workshop on Semantic Evaluation}}. \bibinfo{pages}{1--17}.
\newblock


\bibitem[Morabia et~al\mbox{.}(2019)]%
        {morabia-etal-2019-sedtwik}
\bibfield{author}{\bibinfo{person}{Keval Morabia} {et~al\mbox{.}}} \bibinfo{year}{2019}\natexlab{}.
\newblock \showarticletitle{{SEDTW}ik: Segmentation-based Event Detection from Tweets Using {W}ikipedia}. In \bibinfo{booktitle}{\emph{NACCL workshop}}. \bibinfo{pages}{77--85}.
\newblock


\bibitem[Niu et~al\mbox{.}(2015)]%
        {niu2015topic2vec}
\bibfield{author}{\bibinfo{person}{Liqiang Niu} {et~al\mbox{.}}} \bibinfo{year}{2015}\natexlab{}.
\newblock \showarticletitle{Topic2Vec: Learning distributed representations of topics}. In \bibinfo{booktitle}{\emph{2015 IALP}}. IEEE, \bibinfo{pages}{193--196}.
\newblock


\bibitem[Pellert et~al\mbox{.}(2022)]%
        {pellert2022validating}
\bibfield{author}{\bibinfo{person}{Max Pellert} {et~al\mbox{.}}} \bibinfo{year}{2022}\natexlab{}.
\newblock \showarticletitle{Validating daily social media macroscopes of emotions}.
\newblock \bibinfo{journal}{\emph{Scientific Reports}} \bibinfo{volume}{12}, \bibinfo{number}{1} (\bibinfo{year}{2022}), \bibinfo{pages}{11236}.
\newblock


\bibitem[Penfold and Zhang(2013)]%
        {penfold2013use}
\bibfield{author}{\bibinfo{person}{Robert~B Penfold} {and} \bibinfo{person}{Fang Zhang}.} \bibinfo{year}{2013}\natexlab{}.
\newblock \showarticletitle{Use of interrupted time series analysis in evaluating health care quality improvements}.
\newblock \bibinfo{journal}{\emph{Academic pediatrics}} \bibinfo{volume}{13}, \bibinfo{number}{6} (\bibinfo{year}{2013}), \bibinfo{pages}{S38--S44}.
\newblock


\bibitem[Rao et~al\mbox{.}(2023)]%
        {rao2023tracking}
\bibfield{author}{\bibinfo{person}{Ashwin Rao}, \bibinfo{person}{Rong-Ching Chang}, \bibinfo{person}{Qiankun Zhong}, \bibinfo{person}{Kristina Lerman}, {and} \bibinfo{person}{Magdalena Wojcieszak}.} \bibinfo{year}{2023}\natexlab{}.
\newblock \showarticletitle{Tracking a Year of Polarized Twitter Discourse on Abortion}.
\newblock \bibinfo{journal}{\emph{arXiv preprint arXiv:2311.16831}} (\bibinfo{year}{2023}).
\newblock


\bibitem[Reimers et~al\mbox{.}(2019)]%
        {reimers-2019-sentence-bert}
\bibfield{author}{\bibinfo{person}{Nils Reimers} {et~al\mbox{.}}} \bibinfo{year}{2019}\natexlab{}.
\newblock \showarticletitle{Sentence-BERT: Sentence Embeddings using Siamese BERT-Networks}. In \bibinfo{booktitle}{\emph{In EMNLP-2019}}. \bibinfo{publisher}{ACM}.
\newblock


\bibitem[Rezaei et~al\mbox{.}(2022)]%
        {rezaei2022event}
\bibfield{author}{\bibinfo{person}{Zahra Rezaei} {et~al\mbox{.}}} \bibinfo{year}{2022}\natexlab{}.
\newblock \showarticletitle{Event detection in twitter by deep learning classification and multi label clustering virtual backbone formation}.
\newblock \bibinfo{journal}{\emph{Evolutionary Intelligence}} (\bibinfo{year}{2022}), \bibinfo{pages}{1--15}.
\newblock


\bibitem[Sandoval-Almazan et~al\mbox{.}(2020)]%
        {sandoval2020sentiment}
\bibfield{author}{\bibinfo{person}{Rodrigo Sandoval-Almazan} {et~al\mbox{.}}} \bibinfo{year}{2020}\natexlab{}.
\newblock \showarticletitle{Sentiment analysis of facebook users reacting to political campaign posts}.
\newblock \bibinfo{journal}{\emph{Digital Government: Research and Practice}} \bibinfo{volume}{1}, \bibinfo{number}{2} (\bibinfo{year}{2020}), \bibinfo{pages}{1--13}.
\newblock


\bibitem[Schaffer et~al\mbox{.}(2021)]%
        {schaffer2021interrupted}
\bibfield{author}{\bibinfo{person}{Andrea~L Schaffer}, \bibinfo{person}{Timothy~A Dobbins}, {and} \bibinfo{person}{Sallie-Anne Pearson}.} \bibinfo{year}{2021}\natexlab{}.
\newblock \showarticletitle{Interrupted time series analysis using autoregressive integrated moving average (ARIMA) models: a guide for evaluating large-scale health interventions}.
\newblock \bibinfo{journal}{\emph{BMC medical research methodology}} \bibinfo{volume}{21}, \bibinfo{number}{1} (\bibinfo{year}{2021}), \bibinfo{pages}{1--12}.
\newblock


\bibitem[Schofield(2022)]%
        {Schofield_2022}
\bibfield{author}{\bibinfo{person}{Hugh Schofield}.} \bibinfo{year}{2022}\natexlab{}.
\newblock \bibinfo{title}{French elections: Putin’s war gives Macron boost in presidential race}.
\newblock
\newblock
\urldef\tempurl%
\url{https://www.bbc.com/news/world-europe-60793320}
\showURL{%
\tempurl}


\bibitem[Trager et~al\mbox{.}(2022)]%
        {trager2022moral}
\bibfield{author}{\bibinfo{person}{Jackson Trager} {et~al\mbox{.}}} \bibinfo{year}{2022}\natexlab{}.
\newblock \showarticletitle{The Moral Foundations Reddit Corpus}.
\newblock \bibinfo{journal}{\emph{arXiv preprint arXiv:2208.05545}} (\bibinfo{year}{2022}).
\newblock


\bibitem[Tumasjan et~al\mbox{.}(2010)]%
        {tumasjan2010predicting}
\bibfield{author}{\bibinfo{person}{Andranik Tumasjan} {et~al\mbox{.}}} \bibinfo{year}{2010}\natexlab{}.
\newblock \showarticletitle{Predicting elections with twitter: What 140 characters reveal about political sentiment}. In \bibinfo{booktitle}{\emph{In ICWSM-2010}}, Vol.~\bibinfo{volume}{4}. \bibinfo{pages}{178--185}.
\newblock


\bibitem[vanKleef et~al\mbox{.}(2016)]%
        {vanKleef2016Editorial}
\bibfield{author}{\bibinfo{person}{Gerben~A. vanKleef} {et~al\mbox{.}}} \bibinfo{year}{2016}\natexlab{}.
\newblock \showarticletitle{Editorial: The Social Nature of Emotions}.
\newblock \bibinfo{journal}{\emph{Frontiers in Psychology}}  \bibinfo{volume}{7} (\bibinfo{year}{2016}), \bibinfo{pages}{896}.
\newblock


\bibitem[Weng et~al\mbox{.}(2011)]%
        {weng2011event}
\bibfield{author}{\bibinfo{person}{Jianshu Weng} {et~al\mbox{.}}} \bibinfo{year}{2011}\natexlab{}.
\newblock \showarticletitle{Event detection in twitter}. In \bibinfo{booktitle}{\emph{In ICWSM-2011}}, Vol.~\bibinfo{volume}{5}. \bibinfo{pages}{401--408}.
\newblock


\end{thebibliography}

\appendix
\section{Evaluation of Emotion and Morality Detection} \label{sec:appendix_eval}
We evaluate the effectiveness of emotion and morality detection on a random subset of 850 tweets from the Los Angeles dataset, considering the annotation process is labor and time intensive. We required five educated annotators to go through two training sessions, wherein they annotated 50 random tweets and discussed to improve the agreement on the definitions of emotions and morality. Then each annotator individually annotated all 850 tweets. 
Given the subjectivity of moral and emotional judegments, the Fleiss's $\kappa$ for emotion categories ranges in $0.42 \pm 0.02$. For morality categories, it ranges in $0.30 \pm 0.03$. Similar to prior works \cite{mohammad-etal-2018-semeval,Hoover2020moral}, we have found the $\kappa$ scores of some categories to be low. However, our agreement is still comparable, and on some categories, even better than these prior works. When comparing model performance with the baselines, our methods outperform baselines on most categories (Table~\ref{tab:emot_mf_eval}).

\begin{table}[h]
\caption{Evaluation of Emotion and Morality Detection Methods.}
\centering
\resizebox{0.8\linewidth}{!}{
\begin{tabular}{lcccc|lcccc}
\Xhline{1pt}
\multicolumn{1}{l}{\multirow{2}{*}{\textbf{Emotion}}} & \multirow{2}{*}{\begin{tabular}{c}\textbf{Fleiss’s}\\ $\kappa$ \end{tabular}} & \multicolumn{2}{c}{\textbf{F1-Score}} & \multirow{2}{*}{\textbf{Support}} & \multicolumn{1}{l}{\multirow{2}{*}{\textbf{Morality}}} & \multirow{2}{*}{\begin{tabular}{c}\textbf{Fleiss’s}\\ $\kappa$ \end{tabular}} & \multicolumn{2}{c}{\textbf{F1-Score}} & \multirow{2}{*}{\textbf{Support}} \\ \cline{3-4} \cline{8-9}
\multicolumn{1}{c}{}                                            &                                          & \textbf{Emolex}   & \textbf{Ours}   &                                   & \multicolumn{1}{c}{}                                            &                                          & \textbf{DDR}   & \textbf{Ours}   &                                   \\ \hline
anger                                                           & 0.49                                     & 0.28                & \textbf{0.45}   & 93 & care                                                            & 0.29                                     & \textbf{0.54}       & 0.47            & 63                                \\
anticipation                                                    & 0.32                                     & 0.12                & \textbf{0.40}   & 43 & harm                                                            & 0.28                                     & 0.18                & \textbf{0.36}   & 60                                \\
disgust                                                         & 0.46                                     & 0.33                & \textbf{0.53}   & 116 & fairness                                                        & 0.17                                     & 0.17                & \textbf{0.25}   & 10                                \\
fear                                                            & 0.22                                     & 0.00                & \textbf{0.37}   & 113 & cheating                                                        & 0.28                                     & 0.20                & \textbf{0.43}   & 31                                \\
joy                                                             & 0.48                                     & 0.25                & \textbf{0.37}   & 113 & loyalty                                                         & 0.18                                     & 0.00                & \textbf{0.05}   & 8                                 \\
love                                                            & 0.66                                     & N/A                 & \textbf{0.71}   & 122 & betrayal                                                        & 0.01                                     & 0.00                & 0.00            & 1                                 \\
optimism                                                        & 0.37                                     & 0.15                & \textbf{0.34}   & 58  & authority                                                       & 0.45                                     & 0.00                & \textbf{0.32}   & 24                                \\
pessimism                                                       & 0.26                                     & \textbf{0.20}       & 0.08            & 82  & subversion                                                      & 0.60                                     & 0.15                & \textbf{0.37}   & 78                                \\
sadness                                                         & 0.53                                     & 0.15                & \textbf{0.45}   & 66  & purity                                                          & 0.49                                     & 0.27                & \textbf{0.56}   & 35                                \\
surprise                                                        & 0.51                                     & 0.06                & \textbf{0.44}   & 33  & degradation                                                     & 0.22                                     & 0.21                & \textbf{0.33}   & 23                                \\
trust                                                           & 0.30                                     & 0.00                & \textbf{0.43}   & 22  & & & & &                             \\
\Xhline{1pt}
\end{tabular}}
\label{tab:emot_mf_eval}
\end{table}

\section{Detecting emerging topics after change point}
To determine the new topics relevant to each change point, we compare the top 10 topics before the change point with those after the change point. 
Table~\ref{tab:topic_identification} shows some examples. The newly emerged topics are highlighted in bold. For most reactions, regardless of how small or impactful, the identified topics clearly relate to an offline event (e.g. row 1). For the second example (row 2), the newly emerged topics point us to several different events. However, by examining the tweets belonging to these topics, we found the Black Lives Matter protests was the most predominant event, and other emerging topics such as ``america, vote, trump'' and ``covid, coronavirus, tested'' were related to the protests. Finally, there are also some change points for which we cannot identify meaningful emerging topics (e.g. row 3). We decide whether it is a false positive through manual verification.

\begin{table}[h]
\addtolength{\tabcolsep}{-3.7pt}
\caption{Examples of topics detected before and after several change points.}
\centering
\resizebox{1\linewidth}{!}{
\renewcommand{\arraystretch}{1.15}
\begin{tabular}{c|c|c|c|l|l}
\Xhline{1pt}
\multicolumn{1}{l|}{} & \textbf{\begin{tabular}[c]{@{}c@{}}Change Point\\ Date\end{tabular}} & \textbf{Emotion} & \textbf{Event}                                                        & \multicolumn{1}{c|}{\textbf{Topics Before Change Point}}                                                                                                                                                                                                                                                               & \multicolumn{1}{c}{\textbf{Topics After Change Point}}                                                                                                                                                                                                                                                                                                     \\ \hline
1                     & 2020-05-26                                                           & Betrayal         & \begin{tabular}[c]{@{}c@{}}Black Lives\\ Matter protests\end{tabular} & “president”, “wearing, masks, mask”                                                                                                                                                                                                                                                                                    & \begin{tabular}[c]{@{}l@{}}“\textbf{looting, starts}”,  “\textbf{people, black, white}”,\\  “\textbf{cops, police}”, “president, leader”,\\  “\textbf{fox, news, stand}”, “\textbf{minneapolis, floyd, police}”\end{tabular}                                                                                                                                                                            \\ \hline
2                     & 2020-05-29                                                           & Care             & \begin{tabular}[c]{@{}c@{}}Black Lives\\ Matter protests\end{tabular} & \begin{tabular}[c]{@{}l@{}}“love, sending, thank”, “california, beverly, hills”, \\ “care, feeling, kindness”, “donate, families, wines”, \\ “prayers, pray, praying”, “home, miss, back”, \\ “mask, wearing, wear”, “dog, leash, sorry”, “black”,  \\ “god, bless, blessing”,  “police, officers, cops”\end{tabular}   & \begin{tabular}[c]{@{}l@{}}“love, sending, much”, “donate, donated, black”,\\ “\textbf{america, vote, trump}”, “\textbf{protest, protesting, safe}”,\\  “\textbf{covid, coronavirus, tested}”, “praying, pray, prayers”,\\  “\textbf{peaceful, peach, fighting}”, “\textbf{black, lives, matter}”,\\ “california, los, angeles”, “\textbf{safe, stay, careful}”,\\  “\textbf{city, business, santa, monica}”\end{tabular} \\ \hline
3                     & 2020-06-01                                                           & Pessimism        & Unknown                                                               & \begin{tabular}[c]{@{}l@{}}“sad, heart, people”, “need, rn, scared”, \\ “miss, going, back”\end{tabular}                                                                                                                                                                                                               & \begin{tabular}[c]{@{}l@{}}“sad”, “heart, feel, sleep”, “miss, friend, das”,\\ “heartbreaking, child, right”, “never, wrong, life”\end{tabular}                                                                                                                                                                                                            \\ \hline
\Xhline{1pt}
\end{tabular}}

\label{tab:topic_identification}
\addtolength{\tabcolsep}{3.7pt}
\end{table}

\section{All Events Detected in Different Datasets}
The following tables show all the events detected for each dataset.

\begin{table*}[h]
\caption{Events and the emotional and moral reactions detected in 2020 Los Angeles Data.}
\centering
\resizebox{1\linewidth}{!}{
\renewcommand{\arraystretch}{1.03}
\begin{tabular}{c|l|l|l|l|l|l}
\Xhline{1pt}
& \textbf{Event} & \textbf{Date} & \textbf{Time Window} & \textbf{Peaking Emotion/MF} & \textbf{Declining Emotion/MF} & \textbf{Relevant Topics} \\ \hline
1                     & \begin{tabular}[c]{@{}l@{}}MLK Jr. Day\end{tabular}                                             & 01-20                              & 01-17 to 01-20                                                                                  & joy, loyalty, fairness                                                                                                                                    &                                                                                                         & \begin{tabular}[c]{@{}l@{}}“mlk, king, martin”,\\ “rights, king, justice”\end{tabular}                                                \\ \hline
2                     & \begin{tabular}[c]{@{}l@{}}Trump’s impeachment\\ trial\end{tabular}                                              & 01-20                              & 01-20 to 01-21                                                                                  & betrayal, subversion                                                                                                                                      &                                                                                                         & \begin{tabular}[c]{@{}l@{}}“senate, vote, will”,\\ “treason, traitors, moscow”\end{tabular}                                           \\ \hline
3                    & Earthquake                                                                                                       & 01-20                              & 01-20                                                                                           & fear                                                                                                                                                      & \textbf{}                                                                                               & “earthquake, usgs, km”                                                                                                                \\ \hline
4                     & Kobe Bryant’s death                                                                                              & 01-26                              & 01-26                                                                                           & \begin{tabular}[c]{@{}l@{}}anger, pessimism, sadness,\\ care, harm, purity\end{tabular}                                                                   & joy                                                                                                     & \begin{tabular}[c]{@{}l@{}}“kobe, rest, peace”, \\ “prayers, praying, family”\end{tabular}                                            \\ \hline
5                     & \begin{tabular}[c]{@{}l@{}}Trump’s State of the\\ Union address\end{tabular}                                     & 02-03                              & 02-03                                                                                           & anger, disgust, authority                                                                                                                                 &                                                                                                         & \begin{tabular}[c]{@{}l@{}}“pelosi, nancy, speech”,\\ “pelosi, trump, tear”\end{tabular}                                              \\ \hline
6                     & Valentine’s Day                                                                                                  & 02-14                              & 02-14                                                                                           & love                                                                                                                                                      & anger, disgust                                                                                          & “valentine, valentines, happy”                                                                                                        \\ \hline
7                     & \begin{tabular}[c]{@{}l@{}}CA Pres. primary\end{tabular}                                       & 03-03                              & 03-02 to 03-03                                                                                  & fairness, subversion                                                                                                                                      &                                                                                                         & \begin{tabular}[c]{@{}l@{}}“vote, california”,“biden”\end{tabular}                                              \\ \hline
8                     & COVID-19 pandemic                                                                                                & 03-10                              & 03-09 to 03-11                                                                                  & \begin{tabular}[c]{@{}l@{}}anger, disgust, fear,\\ sadness, care, harm,\\ authority\end{tabular}                                                          & \begin{tabular}[c]{@{}l@{}}anticipation, joy,\\ love, optimism\end{tabular}                             & \begin{tabular}[c]{@{}l@{}}“coronavirus, corona, virus”,\\ “pandemic, virus, administration”,\\ “safe, stay, everyone”\end{tabular} \\ \hline
9                     & \begin{tabular}[c]{@{}l@{}}Earthquake\end{tabular}                       & 04-03                              & 04-03                                                                                           & cheating                                                                                                                                                  &                                                                                                         & “usgs, reports, quake”                                                                                                            \\ \hline
10                     & \begin{tabular}[c]{@{}l@{}}DOJ Drops Flynn Case\end{tabular}                       & 05-07                              & 05-07                                                                                           & cheating                                                                                                                                                  &                                                                                                         & “flynn, cheated, cheating”                                                                                                            \\ \hline
11                    & \begin{tabular}[c]{@{}l@{}}Trump tweeted: ”He got \\caught, OBAMAGATE!”\end{tabular}                             & 05-11                              & 05-11                                                                                           & subversion                                                                                                                                                & \textbf{}                                                                                               & “obama, trump, president”                                                                                                             \\ \hline
12                     & \begin{tabular}[c]{@{}l@{}}BLM protests\end{tabular}                                            & 05-26                              & 05-25 to 05-30                                                                                  & \begin{tabular}[c]{@{}l@{}}anger, disgust, fear, care,\\ harm, fairness, cheating,\\ loyalty, betrayal, authority,\\ subversion, degradation\end{tabular} & \begin{tabular}[c]{@{}l@{}}anticipation, joy,\\ love, optimism\end{tabular}                             & \begin{tabular}[c]{@{}l@{}}“racist, racism, white”,\\ “murder, george, floyd”, \\ “black, injustices, oppresion”\end{tabular}         \\ \hline
13                    & \begin{tabular}[c]{@{}l@{}}Biden criticizes Trump\\ on Russian bounties\end{tabular} & 06-27                              & 06-27                                                                                           & betrayal                                                                                                                                                  & \textbf{}                                                                                               & \begin{tabular}[c]{@{}l@{}}“trump, traitor, must”,\\ “biden, president, treason”\end{tabular}                                         \\ \hline
14                    & \begin{tabular}[c]{@{}l@{}}Dodgers beat Giants\end{tabular}                                              & 07-23                              & 07-23                                                                                           & joy                                                                                                                                                       & \textbf{}                                                                                               & “dodgers, mlb, giants”                                                                                                                \\ \hline                                                                                                        
\Xhline{1pt}
\end{tabular}}

\label{tab:la_reactions}
\end{table*}

\begin{table*}[h]
\caption{Events and the emotional and moral reactions detected in 2022 Abortion data.}
\centering
\resizebox{1\linewidth}{!}{
\renewcommand{\arraystretch}{1.03}
\begin{tabular}{c|l|l|l|l|l|l}
\Xhline{1pt}
& \textbf{Event} & \textbf{Date} & \textbf{Time Window} & \textbf{Peaking Emotion/MF} & \textbf{Declining Emotion/MF} & \textbf{Relevant Topics} \\ \hline
1                     & \begin{tabular}[c]{@{}l@{}}Justice Gorsuch refused \\ to wear mask\end{tabular}                                            & 01-18                              & 01-18                                                                                  & \begin{tabular}[c]{@{}l@{}}betrayal, subversion\end{tabular} &                              & \begin{tabular}[c]{@{}l@{}}“gorsuch, justice, mask”,\\ “refuses, gorsuch, prolife”\end{tabular}         \\ \hline
2                     & \begin{tabular}[c]{@{}l@{}}Taliban crushes Afghan \\ women's rights protest\end{tabular}                                            & 01-18                              & 01-18                                                                                  & \begin{tabular}[c]{@{}l@{}}subversion\end{tabular} & \begin{tabular}[c]{@{}l@{}}love, joy\end{tabular}                             & \begin{tabular}[c]{@{}l@{}}“taliban, women, rights”\end{tabular}         \\ \hline
3                     & \begin{tabular}[c]{@{}l@{}}Jason Miyares filed a motion \\ wanting to overturn Roe v. Wade\end{tabular}                                            & 01-20                              & 01-20                                                                                  & \begin{tabular}[c]{@{}l@{}}authority\end{tabular} & \begin{tabular}[c]{@{}l@{}}optimism\end{tabular}                             & \begin{tabular}[c]{@{}l@{}}“virginia, motion, reversed”,\\ “jason, position, miyares”\end{tabular}         \\ \hline
4                     & \begin{tabular}[c]{@{}l@{}}disturbing video from SCOTUS \\ on Roe v. Wade memorial\end{tabular}                                            & 01-24                              & 01-24 to 01-25                                                                                  & \begin{tabular}[c]{@{}l@{}}disgust, surprise\end{tabular} &                              & \begin{tabular}[c]{@{}l@{}}“wade, roe, memorial”,\\ “shocking, today, video”\end{tabular}         \\ \hline
5                     & \begin{tabular}[c]{@{}l@{}}Val Demings tweeted to support \\ abortion rights\end{tabular}                                            & 02-03                              & 02-03 to 02-04                                                                                  & \begin{tabular}[c]{@{}l@{}}care, joy, love, \\ optimism, trust\end{tabular} & \begin{tabular}[c]{@{}l@{}}subversion, anger, \\ disgust\end{tabular}                             & \begin{tabular}[c]{@{}l@{}}“reproductive, rights, care”,\\ “night, orlando, protecting”\end{tabular}         \\ \hline
6                     & \begin{tabular}[c]{@{}l@{}}Florida banned access to \\ abortions after 15 weeks\end{tabular}                                            & 03-04                              & 03-04                                                                                  & \begin{tabular}[c]{@{}l@{}}harm, cheating, fear\end{tabular} &                              & \begin{tabular}[c]{@{}l@{}}“women, bill, florida”,\\ “senator, movement, crucial”, \\“women, severly, dangerous”\end{tabular}         \\ \hline
7                     & \begin{tabular}[c]{@{}l@{}}Memorial of Death of \\ Breonna Taylor\end{tabular}                                            & 03-08                              & 03-08 to 03-10                                                                                  & \begin{tabular}[c]{@{}l@{}}degradation, care, betrayal\end{tabular} &                              & \begin{tabular}[c]{@{}l@{}}"breonna, tangible, remembering",\\ "praying, seek, care"\end{tabular}         \\ \hline
8                     & \begin{tabular}[c]{@{}l@{}}Marsha Blackburn accused \\ Ketanji Brown Jackson of \\ supporting Griswold v. Connecticut\end{tabular}                                            & 03-21                              & 03-21                                                                                  & \begin{tabular}[c]{@{}l@{}}betrayal\end{tabular} &                              & \begin{tabular}[c]{@{}l@{}}“attacked, griswold, brown”,\\ “ketanji, jackson, prosecuting”\end{tabular}         \\ \hline
9                     & \begin{tabular}[c]{@{}l@{}}Anti-abortionist Mark Robinson \\ says he paid for abortion after \\ impregnating a woman\end{tabular}                                            & 03-23                              & 03-23                                                                                 & \begin{tabular}[c]{@{}l@{}}surprise\end{tabular} &                              & \begin{tabular}[c]{@{}l@{}}“abortion, paid, impregnating”,\\ “surprised, rights, robinson”\end{tabular}         \\ \hline
10                     & \begin{tabular}[c]{@{}l@{}}Oklahoma passed abortion ban \\ punishing people as felons\end{tabular}                                            & 04-05                              & 04-05                                                                                  & \begin{tabular}[c]{@{}l@{}}cheating, subversion\end{tabular} &                              & \begin{tabular}[c]{@{}l@{}}“ban, passed, felons”,\\ “murdered, infant, penalty”, \\“rape, minimum, oklahoma”\end{tabular}         \\ \hline
11                     & \begin{tabular}[c]{@{}l@{}}Republicans forced vote \\ against TitleX\end{tabular}                                            & 04-27                              & 02-26 to 02-28                                                                                  & \begin{tabular}[c]{@{}l@{}}care, degradation, \\subversion\end{tabular} &                              & \begin{tabular}[c]{@{}l@{}}“titlex, screenings, preventative”,\\ “defend, backsliding, titlex”, \\“access, reproductive, program”\end{tabular}         \\ \hline
12                     & \begin{tabular}[c]{@{}l@{}}Leak of SCOTUS ruling \\ to overturn Roe v. Wade\end{tabular}                                            & 05-03                              & 05-02 to 05-03                                                                                  & \begin{tabular}[c]{@{}l@{}}surprise\end{tabular} & \begin{tabular}[c]{@{}l@{}}joy, love, optimism\end{tabular}                             & \begin{tabular}[c]{@{}l@{}}“kept, secrets, confidential”,\\ “overturned, crazy, wild”,\\ “murkowski, rocked, confidence”\end{tabular}         \\ \hline
13                     & \begin{tabular}[c]{@{}l@{}}Memorial of Uvalde \\ school shooting\end{tabular}                                            & 05-24                              & 05-23 to 05-25                                                                                  & \begin{tabular}[c]{@{}l@{}}care, trust, fear\end{tabular} & \begin{tabular}[c]{@{}l@{}}subversion, \\ anticipation\end{tabular}                             & \begin{tabular}[c]{@{}l@{}}“value, life, kids”,\\ “gun, shot, past”\end{tabular}         \\ \hline
14                     & \begin{tabular}[c]{@{}l@{}}Florida synagogue sued \\ new abortion ban\end{tabular}                                            & 06-14                              & 06-14                                                                            & \begin{tabular}[c]{@{}l@{}}purity\end{tabular} &                              & \begin{tabular}[c]{@{}l@{}}“sues, intrusion, prohibits”,\\ “suing, judaism, bible”,\\ “abortion, god, wrong”\end{tabular}         \\ \hline
15                     & \begin{tabular}[c]{@{}l@{}}COVID vaccine protests\end{tabular}                                            & 06-17                              & 06-17                                                                                  & \begin{tabular}[c]{@{}l@{}}subversion\end{tabular} & \                             & \begin{tabular}[c]{@{}l@{}}“vaccine, choice, body”,\\ “biden, executive, action”,\\ “combust, protesters, believe”\end{tabular}         \\ \hline
16                     & \begin{tabular}[c]{@{}l@{}}SCOTUS overturned Roe v. Wade\end{tabular}                                            & 06-24                              & 06-21 to 06-24                                                                                  & \begin{tabular}[c]{@{}l@{}}authority, anger, disgust\end{tabular} & \begin{tabular}[c]{@{}l@{}}care, optimism\end{tabular}                             & \begin{tabular}[c]{@{}l@{}}“roe, happens, sooner”,\\ “patriots, roe, imminent”,\\ “conservative, republican, precedence”,\\ “tyrannized, autocratic, minority”\end{tabular}         \\ \hline
17                     & \begin{tabular}[c]{@{}l@{}}Police arrested House Democrats \\ for protesting Roe v. Wade decision\end{tabular}                                            & 07-19                              & 07-19 to 07-21                                                                                  & \begin{tabular}[c]{@{}l@{}}betrayal, fear, subversion\end{tabular} &                              & \begin{tabular}[c]{@{}l@{}}“police, arrest, dc”,\\ “democratic, inciting, roe, wade”\end{tabular}         \\ \hline
18                     & \begin{tabular}[c]{@{}l@{}}Kansas voted to keep \\ abortion legal\end{tabular}                                            & 08-03                              & 08-02 to 08-03                                                                                  & \begin{tabular}[c]{@{}l@{}}anticipation, trust\end{tabular} &                              & \begin{tabular}[c]{@{}l@{}}“women, health, kansas”,\\ “rebuke, republican, clear”\end{tabular}         \\ \hline
19                     & \begin{tabular}[c]{@{}l@{}}Facebook gave police a \\ teenager's private chats about \\ her abortion\end{tabular}                                            & 08-08                              & 08-08 to 08-09                                                                                  & \begin{tabular}[c]{@{}l@{}}fear, betrayal\end{tabular} &                              & \begin{tabular}[c]{@{}l@{}}“warrants, chats, abortion”,\\ “seize, facebook, phone”,\\ “apps, tracking, surveillance”\end{tabular}         \\ \hline
20                     & \begin{tabular}[c]{@{}l@{}}Texas woman denied an abortion \\ after diagnosis of fatal fetus \\ abnormalities\end{tabular}                                            & 09-28                              & 09-28                                                                                  & \begin{tabular}[c]{@{}l@{}}sadness, surprise, \\subversion\end{tabular} &                              & \begin{tabular}[c]{@{}l@{}}“horrendous, danger, mothers”,\\ “diagnosed, abnormalities, fatal”\end{tabular}         \\ \hline
21                     & \begin{tabular}[c]{@{}l@{}}2022 Midterm election\end{tabular}                                            & 11-08                              & 11-03 to 11-10                                                                                  & \begin{tabular}[c]{@{}l@{}}joy, loyalty, optimism, \\ anticipation, fairness, \\ betrayal, subversion\end{tabular} & \begin{tabular}[c]{@{}l@{}}cheating, fear, anger, \\ disgust, love\end{tabular}                             & \begin{tabular}[c]{@{}l@{}}“choice, body, vote”,\\ “blue, vote, democracy”,\\ “far, media, republicans”\end{tabular}         \\ \hline
22                     & \begin{tabular}[c]{@{}l@{}}Biden Signed 2 orders to protect \\ access to reproductive health care\end{tabular}                                            & 11-19                              & 11-20                                                                                  & \begin{tabular}[c]{@{}l@{}}joy\end{tabular} & \begin{tabular}[c]{@{}l@{}}anger, disgust\end{tabular}                             & \begin{tabular}[c]{@{}l@{}}“protect, care, reproductive”,\\ “executive, signing, biden”\end{tabular}         \\ \hline
23                     & \begin{tabular}[c]{@{}l@{}}Senate stayted blue after election\end{tabular}                                            & 12-07                              & 12-07                                                                                  & \begin{tabular}[c]{@{}l@{}}loyalty, anticipation, \\ optimism\end{tabular} & \begin{tabular}[c]{@{}l@{}}anger, disgust\end{tabular}                             & \begin{tabular}[c]{@{}l@{}}“vote, senate, codify”,\\ “majority, progress, settled”\end{tabular}         \\ \hline
24                     & \begin{tabular}[c]{@{}l@{}}FBI arrested 2 pro-life protesters\end{tabular}                                            & 12-16                              & 12-16                                                                                  & \begin{tabular}[c]{@{}l@{}}harm, fear, \\ sadness, trust\end{tabular} &                              & \begin{tabular}[c]{@{}l@{}}“arrested, violence, facism”,\\ “federally, jesus, individuals”,\\ “advocates, prolifers, attacks”\end{tabular}         \\ \hline

\Xhline{1pt}
\end{tabular}}

\label{tab:abortion_reactions}
\end{table*}

\begin{table*}[h]
\caption{Events and the emotional and moral reactions detected in 2022 French election data.}
\centering
\resizebox{1\linewidth}{!}{
\renewcommand{\arraystretch}{1.03}
\begin{tabular}{c|l|l|l|l|l|l}
\Xhline{1pt}
& \textbf{Event} & \textbf{Date} & \textbf{Time Window} & \textbf{Peaking Emotion/MF} & \textbf{Declining Emotion/MF} & \textbf{Relevant Topics} \\ \hline
1                     & \begin{tabular}[c]{@{}l@{}}Russian-Ukraine war\end{tabular}                                            & 02-23                              & 02-22                                                                                  & \begin{tabular}[c]{@{}l@{}}care, harm\end{tabular} &                              & \begin{tabular}[c]{@{}l@{}}“explosions, thursday, kiev”,\\ “sympathy, shameful, volunteers”, \\ “pendance, donetsk, lugansk”, \\ “emmanuel, macron, condemned”\end{tabular}         \\ \hline
2                     & \begin{tabular}[c]{@{}l@{}}1st round presidential election\end{tabular}                                            & 04-10                              & 04-07 to 04-15                                                                                  & \begin{tabular}[c]{@{}l@{}}subversion, joy, fear\end{tabular} & \begin{tabular}[c]{@{}l@{}}fairness, subversion, \\ anger, disgust\end{tabular}                             & \begin{tabular}[c]{@{}l@{}}“left, macron, turn”,\\ “sir, president, luck”, \\ “macron, facism, maintain”, \\ “will go up, sanctions, moscow”\end{tabular}         \\ \hline
3                     & \begin{tabular}[c]{@{}l@{}}Russian missile cruiser \\ Moskva sank\end{tabular}                                            & 04-14                              & 04-14                                                                                  & \begin{tabular}[c]{@{}l@{}}surprise\end{tabular} &                              & \begin{tabular}[c]{@{}l@{}}“moskva, cruiser, sea”\end{tabular}         \\ \hline
4                     & \begin{tabular}[c]{@{}l@{}}2nd round presidential election\end{tabular}                                            & 04-24                              & 04-21 to 04-26                                                                                  & \begin{tabular}[c]{@{}l@{}}sanctity, fear, \\ sadness, surprise\end{tabular} & \begin{tabular}[c]{@{}l@{}}joy, anger, disgust\end{tabular}                             & \begin{tabular}[c]{@{}l@{}}“liberty, abolition, glory”,\\ “rally, national, respect”, \\ “democracy, emmanuel, result”, \\ “ukraine, conflict, economic”, \\ “macron, price, commodity”\end{tabular}         \\ \hline
5                     & \begin{tabular}[c]{@{}l@{}}Russia cut off natural gas \\ to Poland and Bulgaria\end{tabular}                                            & 04-26                              & 04-26                                                                                  & \begin{tabular}[c]{@{}l@{}}sadness\end{tabular} &                              & \begin{tabular}[c]{@{}l@{}}“gas, bulgaria, poland”, \\ “gas, cut, price”\end{tabular}         \\ \hline
6                     & \begin{tabular}[c]{@{}l@{}}Explosions in Moldova\end{tabular}                                            & 04-26                              & 04-26                                                                                  & \begin{tabular}[c]{@{}l@{}}fear\end{tabular} &                              & \begin{tabular}[c]{@{}l@{}}“transnistria, moldavia, explosions”\end{tabular}         \\ \hline
7                     & \begin{tabular}[c]{@{}l@{}}1st round legislative election\end{tabular}                                            & 06-12                              & 06-12                                                                                  & \begin{tabular}[c]{@{}l@{}}loyalty\end{tabular} & \begin{tabular}[c]{@{}l@{}}anger, disgust\end{tabular}                             & \begin{tabular}[c]{@{}l@{}}“voice, france, legislative”,\\ “struggle, invent, melechon”, \\ “cover, elisabeth, communicate”\end{tabular}         \\ \hline
8                     & \begin{tabular}[c]{@{}l@{}}2nd round legislative election\end{tabular}                                            & 06-19                              & 06-18                                                                                  & \begin{tabular}[c]{@{}l@{}}authority\end{tabular} &                              & \begin{tabular}[c]{@{}l@{}}“nupes, send, vote”,\\ “bravo, congratulations, marine”\end{tabular}         \\ \hline
9                     & \begin{tabular}[c]{@{}l@{}}US supreme court overturned \\ Roe v. Wade\end{tabular}                                            & 06-24                              & 06-24 to 06-25                                                                                  & \begin{tabular}[c]{@{}l@{}}care, harm, cheating, \\ degradation, fear\end{tabular} &                              & \begin{tabular}[c]{@{}l@{}}“abortion, right, your”,\\ “unborn, about, rights”\end{tabular}         \\ \hline
10                     & \begin{tabular}[c]{@{}l@{}}G7 summit\end{tabular}                                            & 06-26                              & 06-24 to 06-25                                                                                  & \begin{tabular}[c]{@{}l@{}}care, fairness, sanctity, \\ fear\end{tabular} &  \begin{tabular}[c]{@{}l@{}}joy\end{tabular}                            & \begin{tabular}[c]{@{}l@{}}“summit, leaders, biden”,\\ “europe, increase, members”, \\ “coordination, strengthen, investments”\end{tabular}         \\ \hline
\Xhline{1pt}
\end{tabular}}

\label{tab:french_reactions}
\end{table*}

\end{document}